\documentclass[aps,prc,preprint,groupedaddress,showpacs,floatfix]{revtex4}

\usepackage{graphicx}
\usepackage{amssymb}
\usepackage{bm}

\newcommand{\be}{\begin{equation}}
\newcommand{\ee}{\end{equation}}
\newcommand{\bel}[1]{\be\label{#1}}
\newcommand{\re}[1]{Eq.~(\ref{#1})}
\newcommand{\ds}{\displaystyle}

\newcommand{\hsp}{\hspace*{1pt}}
\newcommand{\ela}{E_{\rm\hsp lab}}

\begin{document}

\title{
Hydrodynamic modeling of deconfinement phase transition in
heavy-ion collisions at NICA--FAIR energies
}
\author{A.V.~Merdeev}

\affiliation{The Kurchatov Institute, Russian Research Center,
123182 Moscow, Russia}

\author{L.M. Satarov}

\affiliation{Frankfurt Institute for Advanced Studies,~D--60438 Frankfurt
am Main, Germany}

\affiliation{The Kurchatov Institute, Russian Research Center,
123182 Moscow, Russia}

\author{I.N.~Mishustin}

\affiliation{Frankfurt Institute for Advanced Studies,~D--60438 Frankfurt
am Main, Germany}

\affiliation{The Kurchatov Institute, Russian Research Center,
123182 Moscow, Russia}

\begin{abstract}
    We use (3+1) dimensional ideal hydrodynamics to describe the space-time evolution of
    strongly interacting matter created in Au+Au and Pb+Pb collisions. The
    model is applied for the domain of bombarding energies 1--160 AGeV which includes
    future NICA and FAIR experiments. Two equations of state are used: the first one corresponding
    to resonance hadron gas and the second one including the deconfinement phase transition. The initial state
    is represented by two Lorentz-boosted nuclei. Dynamical trajectories of matter in the central
    box of the system are analyzed. They can be well represented by a fast shock--wave compression
    followed by a relatively slow isentropic expansion.  The parameters of collective flows and hadronic
    spectra are calculated under assumption of the isochronous freeze-out. It is shown that the deconfinement
    phase transition leads to broadening of proton rapidity distributions, increase of elliptic flows
    and formation of the directed antiflow in the central rapidity region. These effects are most pronounced
    at bombarding energies around 10~AGeV, when the system spends the longest time in the mixed phase.
    From the comparison with three--fluid calculations we conclude that the transparency
    effects are not so important in central collisions at NICA--FAIR energies (below 30 AGeV).
\end{abstract}


\maketitle

\section{Introduction}

Relativistic hydrodynamics is a very powerful tool to study high--energy nuclear collisions.
Especially useful is its sensitivity to the ''equation of state'' (EoS)
of strongly interacting matter and, in particular, to its phase diagram. In fact, extracting this EoS
the main goal of heavy--ion experiments. Theoretical studies of the EoS are still not very conclusive.
First principle QCD calculations on the lattice~\cite{Baz09,Aok09} give reliable results
only at small baryon chemical potentials. In this case
a cross-over transition from hadronic to quark--gluon degrees of freedom is predicted at temperatures
\mbox{$T\sim 17\hsp 0$ MeV}. Some signatures of a baryon--free quark-gluon plasma with low viscosity
have been already found~\cite{Ada05} in RHIC experiments with c.m. bombarding
energies \mbox{$\sqrt{s_{NN}}=60-200$ GeV}. These signatures may be even stronger at LHC energies
(\mbox{$\sqrt{s_{NN}}\sim 5$ TeV}). On the other hand, many phenomenological models predict
that a strong first order phase transition may occur in compressed baryon--rich
matter~\cite{Alf98,Sca01,Ste04}. Presumably, such matter is created at lower (SPS, AGS) energies.
A~more detailed data should be obtained in the low--energy runs at RHIC~\cite{Rit06} as well as in
future NICA~\cite{Nic11} and FAIR~\cite{Fai06} experiments.

A large amount of experimental data for nuclear collisions at AGS, SPS and RHIC energies
have been successfully described by different versions of the hydrodynamic model.
The first model of this kind has been proposed by Landau more that 50 years ago~\cite{Lan53}.
Unless stated otherwise, below we are dealing with perfect fluids, i.e. we neglect possible
dissipative terms, associated with viscosity, heat conductivity as well as
chemical non-equilibrium effects. In other words, it is assumed that deviations from local
equilibrium are small starting already from early stages of a nuclear collision.
One can roughly divide the existing versions of
ideal hydrodynamics into two classes. The first class includes the models which apply fluid
dynamical simulations from the very beginning i.e. starting from cold equilibrium nuclei.
The attractive feature of such an approach is that no additional parameters are needed to characterize
the initial state of the reaction. The models of the second class introduce an~excited
and compressed initial state -- a locally equilibrated ''fireball''. It is believed that such
fireball is formed at an early non-equilibrium stage of the collision. The disadvantage of this
approach is a large freedom in choosing geometrical and fluid-dynamical parameters
of the initial state. Due to this reason the predictive power of second-class models is greatly reduced,
especially when studying sensitivity of the results to the EoS. Up to now many versions of
the fireball--based hydrodynamic model were developed ranging from simplified
(1+1)--~\cite{Bjo83,Mis83,Bla87,Esk98,Moh03,Sat07} and
\mbox{(2+1)--dimensional} models~\cite{Bla87,Kol99,Bas00,Per00,Tea01} to more sophisticated
(3+1)--dimensional ones~\cite{Sol97,Non00,Agu02,Hir02,Boz09a,Sch10}.
Recent calculations with inclusion of dissipative terms~\cite{Son08,Luz08} show that data at RHIC
energies can be reproduced with rather low viscosity coefficients.

Historically, early 3D models of relativistic nuclear collisions~\cite{Ams75,Sto79,Ros81,Bra94,Ris95a,Pae01}
used cold Lorentz-contracted nuclei in the initial state. It is believed that such models are good enough
up too bombarding energies of about 10 AGeV. Due to the projectile--target
transparency, they become less and less justified with increasing bombarding energy. To take into account
this effect, generalized multi-fluid models have been constructed in
Refs.~\cite{Ams78,Cla86,Mis88,Kat93,Bra00,Iva06}. The most important ingredients of such models, the
inter-fluid coupling terms, are rather uncertain and usually parameterized phenomenologically.

It is clear that hydrodynamical approach can not be directly applied to late
stages of a heavy--ion reaction when binary collisions of particles become too rare to maintain
the local thermodynamic equilibrium. A standard way~\cite{Lan53} to circumvent this difficulty is
to introduce a so called ''freeze--out'' criterion to stop the hydrodynamical description. Often
it is postulated that collisionless expansion of particles starts at some isothermal hypersurface~\cite{Coo74}.
Unfortunately, this approximation is rather crude~\cite{Bug96} and even in contradiction with experimental
data (see e.g.~\cite{Hun98}). A more consistent procedure has been
suggested~\mbox{\cite{Bas00,Tea01,Non07,Hir08,Pet08}} within a hybrid ''hydro--cascade'' model.
In this scheme hydrodynamics is used for generating coordinates and momenta  of hadrons at
some intermediate stage of the reaction. These characteristics are used for transport simulations
of later stages.

Below we formulate a version of the ideal (3+1)--dimensional hydrodynamics suitable
for the domain of AGS, NICA, FAIR and SPS energies. This model belongs to the first class
and uses a new EoS~\cite{Sat09} with the deconfinement and liquid--gas phase transitions.
In our calculations we perform a detailed analysis of the dynamics of nuclear collisions
at various energies giving a particular attention to macroscopic characteristics, i.e. collective flows,
life--time and volume of quark--gluon and mixed phases. It is shown that maximal energy and baryon densities
predicted by our fluid--dynamical simulations agree well with 1D shock wave calculations. A special
analysis is carried out to evaluate the importance of transparency effects. This is made by comparing
our results with predictions of the three--fluid model of Ref.~\cite{Iva06}.
To calculate momentum spectra as well as directed and elliptic flows of secondary particles,
we apply an approximation of isochronous freeze-out. In these calculations we investigate the sensitivity of
observables to finite size of hadrons by introducing the excluded volume corrections.

The paper is organized as follows: our hydrodynamic model is formulated in Sec.~II. The results of numerical
calculations for central Au+Au collisions at FAIR and SPS energies are given
in Sec.~III . Here we also compare the predictions of the one-- and three--fluid models.
In Sec.~IV we discuss a space-time picture of non-central Au+Au collisions. Particle spectra
and parameters of transverse collective flows are considered in Sec.~V and VI.
Our conclusions are presented in Sec.~VII.

\section{Formulation of the model}

\subsection{Equations of ideal fluid dynamics}

    Below we study the evolution of highly excited, and possibly deconfined, strongly--interacting
    matter produced in ultrarelativistic heavy-ion collisions. It is assumed that this evolution can
    be described by the equations of ideal relativistic hydrodynamics~\cite{Lan87}. These equations
    represent local conservation laws of the 4-momentum and baryon charge
\begin{eqnarray}
            &&\partial_\nu T^{\mu\nu}=0\hsp,
        \label{emcon}\\
            &&\partial_\mu N^\mu=0\hsp.
        \label{chcon}
\end{eqnarray}
    In the limit of small dissipation the baryon 4--current $N^\mu$ and the energy-momentum
    tensor~$T^{\mu\nu}$ can be expressed as ($\hbar=c=1$)
        \bel{bden}
            N^{\mu}=n u^{\mu},
        \ee
        \bel{emten}
            T^{\mu\nu}=(\varepsilon+P)\hsp u^\mu u^\nu -P\hsp g^{\hsp\mu\nu},
        \ee
    where $\varepsilon, n$ and $P$ are the rest--frame energy density, the net baryon density and
    pressure of the fluid,
    $u^\mu = \gamma\hsp (1,\bm{v})^{\hsp\mu}$ is its collective 4--velocity, $\gamma = (1 - \bm{v}^2)^{-1/2}$,
    $g^{\hsp\mu\nu}=\textrm{diag}\,(+,-,-,-)$ is the metric tensor. Here and below we denote
    by $\bm{v}$ the fluid 3--velocity.

    One can rewrite Eqs.~(\ref{emcon})--(\ref{chcon}) in Cartesian coordinates $(t,\bm{r})$
    as follows
  \begin{eqnarray}
            &&\partial_t {E}\:+\nabla (\bm{v} E)=-\nabla (\bm{v} P)\hsp,
        \label{ender}\\
            &&\partial_t {\bm{M}}\:+\nabla (\bm{v} \bm{M})=-\nabla P\hsp,
        \label{mder}\\
            &&\partial_t N\:+\nabla (\bm{v} N)=0\hsp,
        \label{nder}
\end{eqnarray}
    where \mbox{$E=T^{\hsp 00}\hspace*{-1pt},\,  M^{\hsp i}=T^{\hsp 0i}$} and \mbox{$N=N^{\hspace*{.5pt}0}$}
    are the energy density,
    the 3-momentum density and the baryon density in the lab frame.

    The relations between hydrodynamical variables in the lab and local rest frames can be written as
  \begin{eqnarray}
             &&E=\gamma^2\hsp (\varepsilon+v^2 P),
        \label{elden}\\
            &&\bm{M} =\gamma^2\hsp (\varepsilon+P)\,\bm{v},
        \label{mlden}\\
            &&N=\gamma\hsp n\hsp.
        \label{blden}
\end{eqnarray}
    To solve Eqs.~(\ref{ender})--(\ref{blden}) it is necessary to specify the
    EoS $P=P(n,\varepsilon)$ of the fluid and initial conditions.

\subsection{Equation of state}\label{EoS}

        \begin{figure*}[htb!]
        \centerline{\includegraphics[width=0.8\textwidth]{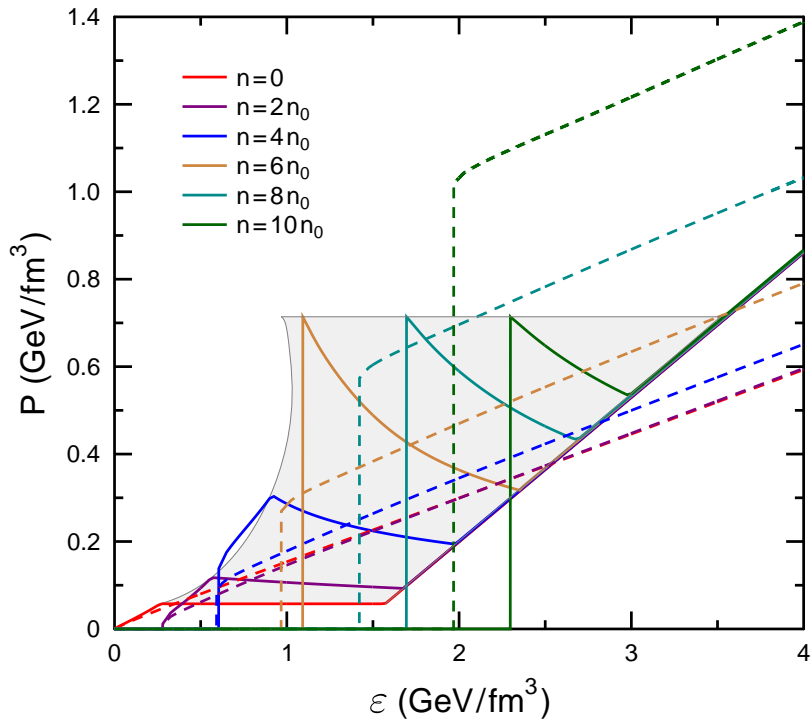}}
        \caption{(Color online)
        Pressure as a function of energy density at different baryon densities, scaled by
        saturation nuclear density \mbox{$n_0=0.15\,{\rm fm}^{-3}$}.
        Solid and dashed lines correspond to EoS--PT and EoS--HG, respectively.
        The MP region of the deconfinement phase transition is shown by shading.
        }
        \label{fig1}
        \end{figure*}
        \begin{figure*}[htb!]
        \centerline{\includegraphics[width=0.8\textwidth]{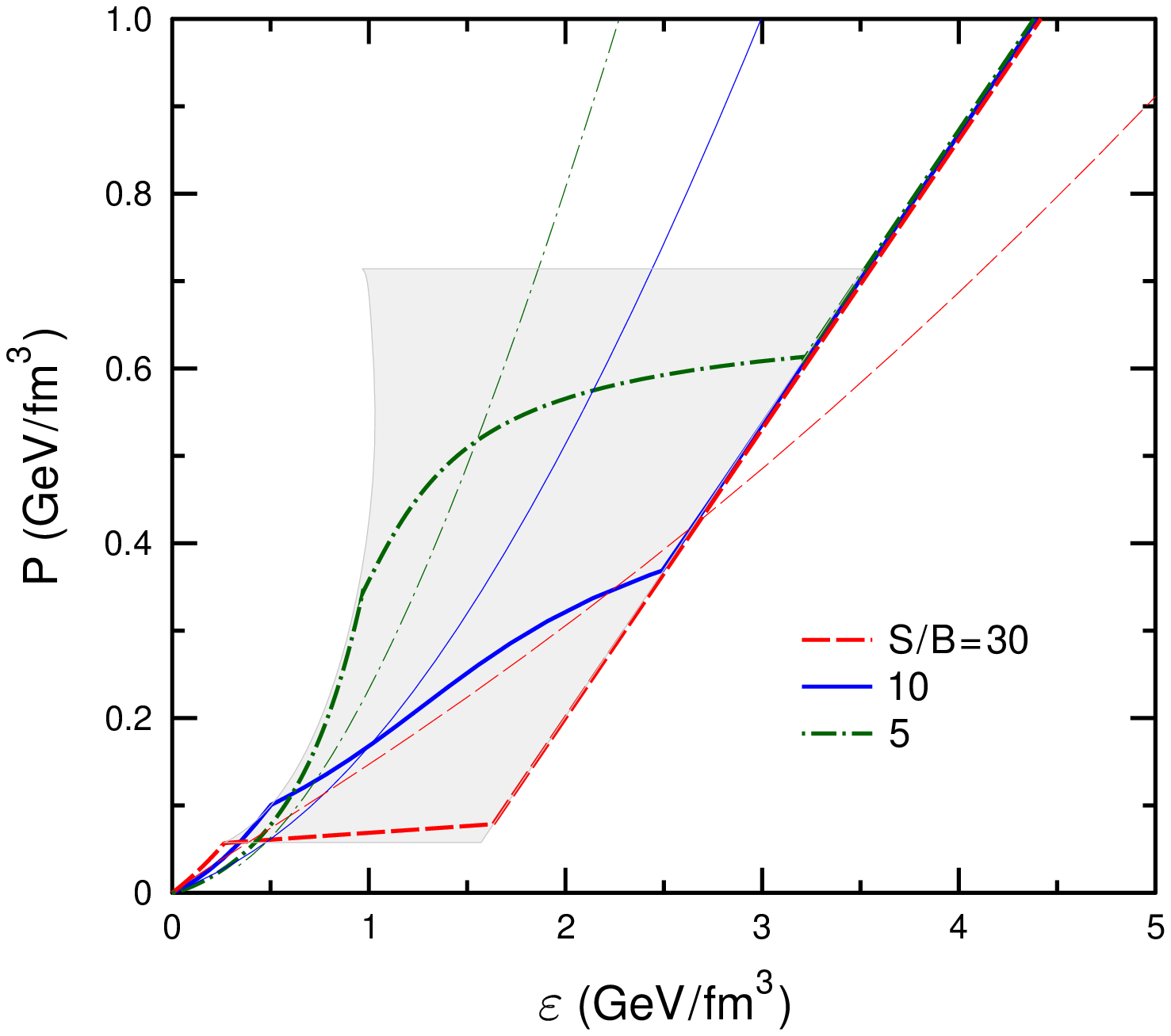}}
        \caption{(Color online)
        Adiabatic trajectories in the $\varepsilon-P$ plane.
        Thick and thin lines correspond to EoS--PT and EoS--HG, respectively.
        Shading shows the MP region of the deconfinement phase transition.
        }
        \label{fig2}
        \end{figure*}

    In our calculations we use the EoS of strongly interacting matter
    with the first order deconfinement phase transition (EoS--PT)~\cite{Sat09}.
    The hadronic phase (HP) is described as the hadron resonance gas
    with inclusion of known hadrons with masses up to 2 GeV.
    Finite size effects are taken into account by introducing the excluded volume
    corrections. The same excluded volume parameter \mbox{$v_e=1$ fm$^3$} is used
    for all hadronic species. A Skyrme--like mean field for baryons
    \bel{smf}
    U(n)=-\alpha\hsp n+\beta\hsp n^{7/6}
    \ee
    is added to guarantee that the hadronic matter has correct saturation point and a
    liquid-gas phase transition~\footnote
    {
    In this paper we do not study effects connected with the liquid-gas fragmentation of relatively cold
    hadronic matter. Below, when speaking about the phase transition we have in mind
    the deconfinement--hadronization transition.
    }.
    The deconfined quark--gluon phase (QGP) is described by the bag model EoS with lowest-order perturbative
    corrections. The phase transition boundaries and characteristics of the mixed phase (MP) are found
    by using the Gibbs conditions with the strangeness neutrality constraint
    (for details, see Ref.~\cite{Sat09}).
    The domains of MP states in different thermodynamic variables are shown in Figs.~\ref{fig1}--\ref{fig2},
    \mbox{\ref{fig6}--\ref{fig7}}, \mbox{\ref{fig9}--\ref{fig10}}.
    To probe sensitivity to the EoS, we have also performed calculations with the EoS of ideal hadron gas
    (EoS--HG). In this case we disregard the excluded volume effects assuming that \mbox{$v_e=0$}.
    To provide stability of initial nuclei we also introduce the Skyrme-like mean field. The parameters of
    EoS--PT and EoS--HG are given in~\cite{Sat09}.

    The graphic representation of $P(n,\varepsilon)$ for both EoS\hspace*{0.5pt}s is given in Figs.~\ref{fig1}--\ref{fig2}
    \footnote{
    Note, that jumps of pressure curves in Fig.~\ref{fig1} correspond to the zero temperature boundary of
    physical states in the $n-\varepsilon$ plane.
    }.
    One can see that values of pressure predicted by these two \mbox{EoS\hspace*{0.5pt}s} significantly differ
    at large enough~$\varepsilon$ and $n$.
    An important information about the dynamics of matter in the heavy-ion
    collision can be drawn from Fig.~\ref{fig2} which shows adiabatic trajectories, i.e. lines of constant
    entropy per baryon, $\sigma=s/n$, where $s$ is the entropy density. As well known~\cite{Lan87},
    this quantity
    is conserved along the trajectories of fluid elements in the ideal hydrodynamics (in absence of
    shock waves).
    Our calculations show that larger values of $\sigma$ in central heavy--ion collisions
    are achieved at larger bombarding energies $\ela$. The slopes of adiabates in Fig.~\ref{fig2}
    give the sound velocities squared~\cite{Lan87}
      \bel{svel}
      c_s^{\,2}=\left(\frac{\partial P}{\partial\hsp\varepsilon}\right)_{\hspace*{-2pt}\sigma}.
      \ee
   From from Fig.~\ref{fig2} one can clearly see that $c_s$ reach minimal values (''softest points'')
   in the MP region. This should essentially influence the expansion dynamics. Indeed,
   an instantaneous acceleration of a fluid element is proportional to
   the gradient of pressure $\nabla\hspace*{-1.5pt}P=c_s^{\,2}\hsp\nabla\varepsilon$\,.
   At given $\nabla\varepsilon$ one may
   expect smaller accelerations of matter for ''softer'' EoS\hspace*{0.5pt}s, which have smaller values of sound velocity.
    From Fig.~\ref{fig2} one can see that although generally the EoS with the phase transition is softer than
    the EoS of pure hadron gas, in some regions of thermodynamic parameters the EoS--PT has sound
    velocities of the same order or even larger than those in the EoS--HG. Such a situation may
    occur in the MP region for intermediate values of specific entropy. This anomaly is expected in the
    NICA--FAIR energy domain, where our calculations predict a non-monotonous behavior of collective
    flows and particle spectra as functions of bombarding energy (see below).

\subsection{Initial conditions}

    We start our hydrodynamical simulation from the stage when two cold nuclei approach each
    other. Unless stated otherwise, we consider collisions of gold nuclei ($Z=79,A=197$) which have
    the Woods-Saxon density distribution in their rest frame,
        \bel{dinit}
            n\hsp(\bm r,t=0)=\frac{n_0}{1+\exp{\left(\frac{\ds |\bm r| - R_{\hsp 0}}{\ds a}\right)}}\,,
        \ee
    with the parameters \mbox{$n_0=0.15$ fm$^{-3}$}, \mbox{$a=0.5$ fm}, \mbox{$R_{\hsp 0}=6.7$ fm}.
    We assume that the rest frame initial energy density of each nucleus, $\varepsilon\hsp(\bm r,t=0)$
    is given by the r.h.s.~of~\re{dinit}
    with the re\-pla\-ce\-ment of~$n_0$ by \mbox{$\varepsilon_0=\mu_0\hsp n_0$} where
    \mbox{$\mu_0=0.923$ GeV} is the baryon chemical potential of equilibrium nuclear matter.
    Due to the stabilizing effect of the Skyrme-like mean field, the initial nuclei may stay
    in equilibrium with vacuum at $P=0$\,. They propagate without distortion until their density
    distributions essentially overlap.

    The c.m.~collective velocity of nuclei at \mbox{$t=0$} is taken as
    \mbox{$v_0=\sqrt{\ela/(2\hsp m_N+\ela)}$}\hsp,
    where \mbox{$m_N=0.939$ GeV} is the nucleon mass and $\ela$ is the projectile bombarding energy (per
    nucleon) in the lab frame. Below the beam axis is denoted by $z$ and the $x$--axis is chosen along
    the impact parameter vector $\bm{b}$.
    In this case the reaction and transverse planes correspond to \mbox{$y=0$} and \mbox{$z=0$}, respectively.
    The initial longitudinal distance between the target and projectile centers in the c.m. frame is chosen as
        \bel{indis}
            z_{\hsp t}(0)-z_p(0)=2\hsp (R_{\hsp 0}+6.9\hsp a)/\gamma_0\,,
        \ee
    where $\gamma_0=(1-v_0^2)^{-1/2}$ is the Lorentz factor of colliding nuclei\,\footnote
    {
    From Eqs.~(\ref{dinit}), (\ref{indis}) one can see that in the case of a central collision
    the rest frame density of each nucleus at \mbox{$x=y=z=ct=0$}  equals approximately $10^{-3}n_0$.
    }.

\subsection{Numerical scheme} \label{snums}

The numerical solution of fluid--dynamical equations (\ref{ender})--(\ref{nder}) was obtained by using the
flux-corrected transport algorithm~\cite{Bor73,Ris95b}.
 \begin{figure*}[htb!]
        \centerline{\includegraphics[width=0.8\textwidth]{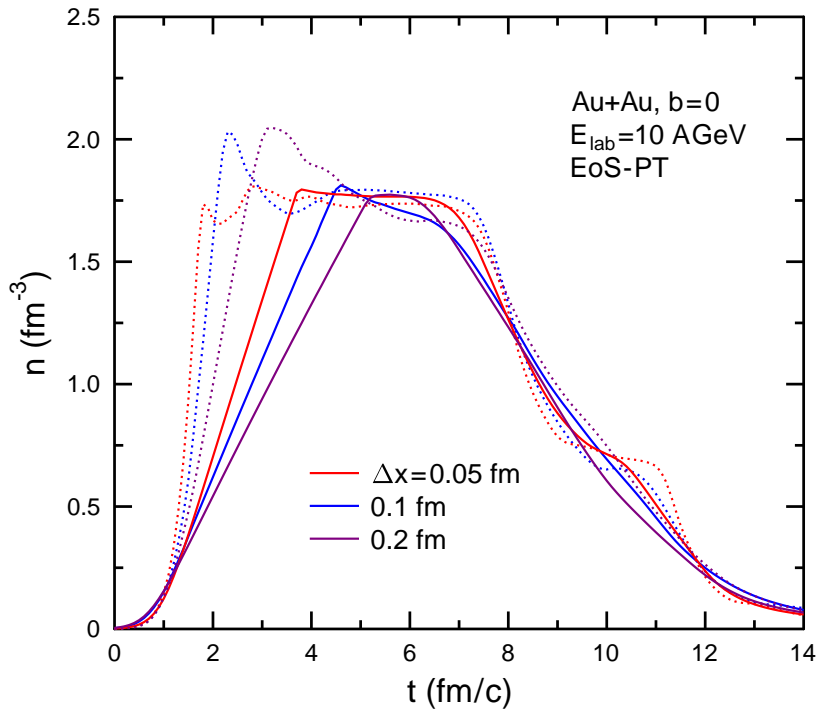}}
        \caption{(Color online)
        The rest--frame baryon density $n$ as a function of time in a central ($b=0$) Au+Au collision
        at $\ela=10$ AGeV. Dotted lines show density in a central cell, while solid ones represent
        density values averaged over the central box. Different colors correspond to calculations
        with different $\Delta x$. All results are obtained for EoS--PT.
        }
        \label{fig3}
        \end{figure*}
All calculations are performed in the c.m. frame.
Unless otherwise stated, we use a cubic Eulerian grid
with the cell size \mbox{$\Delta\hsp x=0.1$ fm} along each direction. Typically we take the numbers of grid points
equal to $500\times 300\times 400$ in $x,y,z$ directions and choose the time step \mbox{$\Delta\hsp t=0.01$
fm/$c$}. In our simulations we use linear interpolations of tables $P(n,\varepsilon)$ prepared with fixed
steps in $n$ and $\epsilon$. We have checked that our numerical scheme conserves the total baryon number and energy with
relative accuracy better than 1\%. The program code explicitly uses the symmetry of the collision process with
respect to the reaction plane~\mbox{$y=0$}.
This reduces the computational memory by a factor of two. A typical run for the 10 AGeV Au+Au collision at
$b=7$ fm requires about 6 Gb of memory and approximately 6 hours of CPU time on a machine with
the 2.1 GHz AMD Opteron processor.

Below we often analyze time evolution of fluid-dynamical quantities averaged over a ''central box''
around the symmetry point \mbox{$x=y=z=0$}. As a rule, we use the box with dimensions~\footnote
    {
    At \mbox{$\ela\ge 80$ AGeV} we use energy--independent
    sizes of the central box: \mbox{$|x|,|y|<1$ fm}, \mbox{$|z|<0.2$ fm}. In Fig.~\ref{fig8} a larger
    box is used with sizes increased by a factor of 2 as compared to \re{cbox}.
    }
    \bel{cbox}
    |x|,\,|y|,\,\gamma_0|z|\,<\,1\,{\rm fm}\hsp.
    \ee
For comparison Fig.~\ref{fig3} shows the time dependence of the baryon density in the central grid cell
and in the central box for the case of a central Au+Au collision at  \mbox{$\ela=10$ AGeV}. One can
see that using rough grids leads
to oscillations of the energy density which are of numerical origin. They are especially visible at
early stages when a shock--like compression of matter takes place (see Sec.~IIIB). Averaging over
neighboring cells partly remove these oscillations.

\subsection{Particle spectra and parameters of collective flows}\label{specf}

To calculate hadronic momentum distributions we apply the approximation
of instantaneous freeze-out: it is assumed that a sudden transition
from the local equilibrium to collisionless propagation of particles
takes place at some space--time hypersurface~$\sigma_\mu$~\footnote
{
It is also postulated that presence of already frozen fluid elements does not influence
significantly the dynamics of remaining parts of the system.
}.
Within this approximation one can use the standard Cooper--Frye formula~\cite{Coo74}
for the invariant momentum distribution of the hadronic species $i$
\bel{spec}
E\hsp\frac{d^{\hsp 3}N_i}{\hspace*{-4pt}d^{\hsp 3}p}=\frac{d^{\hsp 3}N_i}{dy\hsp d^{\hsp 2}p_{\hsp T}}=
\frac{g_i}{(2\pi)^{\hsp 3}}
\int d\sigma_\mu\hsp p^{\hsp\mu}\left\{\exp\left(\frac{p_\nu u^\nu-\mu_i}
{T}\right)\pm 1\right\}^{-1},
\ee
where \mbox{$p^{\hsp\mu}=(E,\bm{p})^\mu$} is the 4--momentum of the particle,
$y$ and $\bm{p}_{\hsp T}$ are, respectively, its longitudinal rapidity and
transverse momentum, $T$ denotes the local temperature, $g_i$ is the statistical weight of the
hadron species~$i$.

Plus and minus in \re{spec} correspond, respectively, to fermions and bosons.
Using conditions of chemical equilibrium one can express the particle's
chemical potential~$\mu_i$ through the baryon~($\mu$) and strange ($\mu_S$)
chemical potentials as follows~\cite{Sat09,Lan80}
\bel{cceq}
\mu_i=B_i\mu+S_i\mu_S\,,
\ee
where $B_i$ and $S_i$ are, respectively, the baryon and strangeness number
of species $i$\,.

In calculating spectra of baryons ($B_i=1$) we take into account the mean field effects.
In this case we use Eqs.~(\ref{spec})--(\ref{cceq}) with the replacement of the baryon
chemical potential~$\mu$ by its ''kinetic'' part $\mu_K=\mu-U(n)$ where $U(n)$ is the
mean field potential introduced in~\re{smf}. As indicated above, our calculations with
the deconfinement phase transition use the EoS with inclusion of finite volume corrections.
In accordance with Ref.~\cite{Sat09}, we introduce these corrections by the replacement
$\mu_i\to\mu_i-v_iP_K$ where $P_K$ is the kinetic part~\cite{Sat09} of the total pressure
and $v_i$ is the excluded volume of $i$-th hadron. Unless stated otherwise,
particle spectra and collective flows in the case of EoS--PT are calculated by using the
parameter $v_i=1\,\textrm{fm}^3$ for all hadronic species.

For our qualitative analysis we choose the simplest condition of isochronous
freeze--out (\mbox{$t=t_{\rm fr}={\rm const}$}).
Then $d\sigma_\mu=d^{\hsp 3}x\hsp\delta_{\mu,0}$ and \re{spec} gives the
hadronic momentum distribution at a fixed time~$t$ in the c.m. frame.
In addition to contributions of ''thermal'' nucleons and pions, which are calculated
directly by using~\re{spec}, we also take into account resonance decays, e.g.
\mbox{$\Delta\to N\pi$}, \mbox{$\rho\to 2\pi$}. Below we assume zero widths of resonances
and apply the following formula~\cite{Sat07} for the spectrum
of hadrons $i$ from two--body decays \mbox{$R\to i+\hspace*{-1pt}X$}~\footnote
{
The decay channels with more than two final hadrons are included approximately.
Namely, we consider the decay into $n+1$ hadrons
\mbox{$R\to i+h_1+\ldots +h_n (n>1)$}
as a reaction $R\to i+X$, where $X$ is a hadronic cluster with mass equal to the
total mass of hadrons \mbox{$h_1,\ldots h_n$} at rest.
}
\bel{stbd}
E\hsp\frac{d^{\hsp 3}N_{R\hsp\to\hsp i}}{\hspace*{-1.9em}d^{\hsp 3}p}=\frac{b_i}{4\pi q_0}
\int d^{\hsp 3}p_R\,\frac{d^{\hsp 3}N_R}{\hspace*{-4pt}d^{\hsp 3}p_R}
\,\delta\left(\frac{p\hsp p_R}{m_R}-E_0\right).
\ee
Here $m_R$ and $p_R$ are the mass and 4--momentum of the resonance $R$ ($R=\Delta,\rho\ldots$)\hsp,
$E_0$~and~$q_0$ are the energy and 3--momentum of the hadron $i$ in the rest frame of $R$,
$b_i$ is the branching ratio of the decay channel \mbox{$R\to i+\hspace*{-1pt}X$}.
Equation~(\ref{spec}) is used to calculate spectrum of resonances (in the Boltzmann approximation).
All known hadronic resonances with masses up to 2 GeV are taken into account.
The statistical weights,
branching ratios and masses of these resonances are taken from~\cite{Nak10}.

Transverse collective flows of matter created in heavy--ion collisions,
are rather sensitive to its EoS. Much attention in recent years is given to the parameter of
elliptic flow $v_2$~\cite{Oll92}. The 2+1 fluid dynamical calculations at SPS and
RHIC energies~\cite{Tea01,Kol00} show a significant sensitivity of this quantity
to the deconfinement phase transition. The elliptic flow of $i$--th hadrons
($i=\pi,N\ldots$) is usually determined as
\bel{eqv2}
v_2^{(i)} (y)=
\frac{\ds\int d^{\hsp 2} p_{\,T} \cos{(2\hsp\phi)}\,
E\hsp d^{\hsp 3}\hspace*{-1pt}N_i/d^{\hsp 3}p}{\ds\int d^{\hsp 2} p_{\,T}
E\hsp d^{\hsp 3}\hspace*{-1pt}N_i/d^{\hsp 3}p}\,,
\ee
where \mbox{$\phi=\arccos{(p_{\,x}/p_{\,T})}$} is the azimuthal angle of $\bm{p}_{\,T}$
with respect to the reaction plane. Here $d^{\hsp 3}\hspace*{-1pt}N_i/d^{\hsp 3}p$
is the momentum distribution of $i$--th hadrons with inclusion of
resonance decays. Unfortunately, the elliptic flow is sensitive~\cite{Hir08,Iva09a,Pet10} not only to the
EoS, but also to dissipative and freeze--out effects. Below we also calculate the directed flow
parameter~$v_1^{(i)}(y)$. It is defined by the r.h.s. of \mbox{\re{eqv2}} with
the replacement $\cos{(2\hsp\phi)}\to\cos{\hsp\phi}$. It is clear that \mbox{$v_1, v_2\to 0$} for
purely central collisions (\mbox{$b\to 0$})\hsp.

To discuss qualitatively possible differences between the EoS--PT and EoS--HG at different bombarding energies, below we
calculate a so--called momentum anisotropy parameter~$\epsilon_p$ which characterizes the collective flow asymmetry
in the transverse plane. Following Ref.~\cite{Boz09} we define this quantity as
\bel{anism}
\epsilon_p=\frac{\ds\int dx\hsp dy\left(T^{\hsp xx}-T^{\hsp yy}\right)}
{\ds\int dx\hsp dy\left(T^{\hsp xx}+T^{\hsp yy}\right)}\,,
\ee
where $T^{\hsp xx}, T^{\hsp yy}$ are the components of momentum density in the transverse plane
\mbox{$z=0$}\,\footnote
{\label{expl}
Experimentally one can not fix longitudinal coordinates of particles in a central slice. Instead, those
particles are usually observed, whose c.m. rapidities~$y_i$ fall in some window~\mbox{$|y_i|<y_*$}.
In the Bjorken scaling picture, such a window corresponds to an expanding layer
\mbox{$|z_i|<\Delta z=\tanh{y_*}\hsp t$}\,. To take this into account, we perform also calculations
using a modified definition of the momentum anisotropy, with additional
integration over $|z|<\Delta z$ in the numerator and denominator
of~\re{anism}. This modification is applied only for $t>R/\gamma_0$.
}.
In Ref.~\cite{Kol99} the approximate relation \mbox{$\epsilon_p\simeq 2\hsp v^{(\pi)}_2(0)$}
has been obtained from (2+1)--hydrodynamical simulations of Pb+Pb collisions at SPS and RHIC energies.

\section{Evolution of fluid--dynamic quantities in central collisions}

\subsection{Comparison of results for EOS--PT and EOS--HG}

        \begin{figure*}[htb!]
        \centerline{\includegraphics[width=\textwidth]{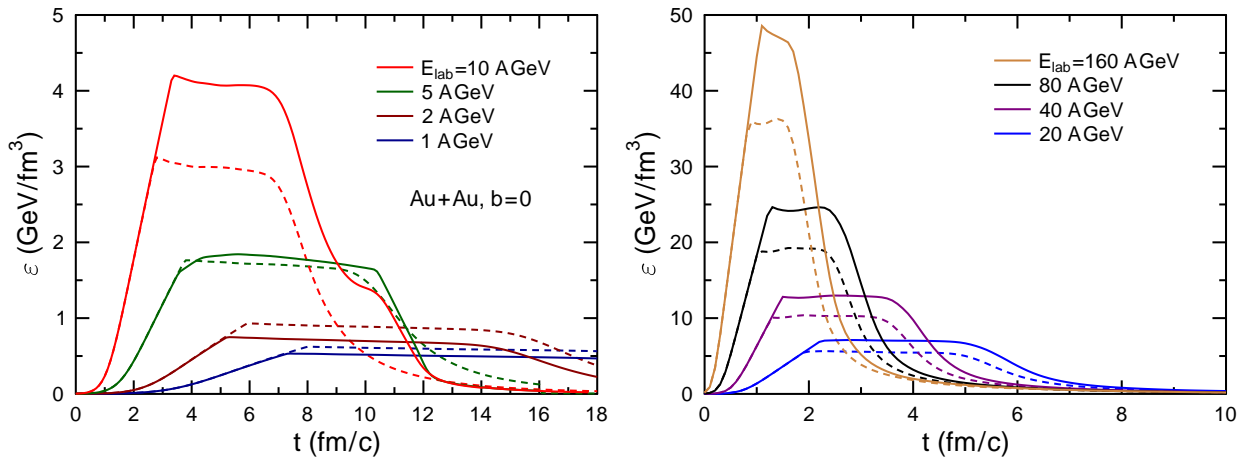}}
\caption{(Color online)
Time evolution of the energy density in the central box of a central Au+Au collision
for various bombarding energies. Solid and dashed lines correspond to EoS--PT and \mbox{EoS--HG},
respectively.
}
        \label{fig4}
        \end{figure*}
\begin{figure*}[htb!]
        \vspace*{2mm}
        \centerline{\includegraphics[width=\textwidth]{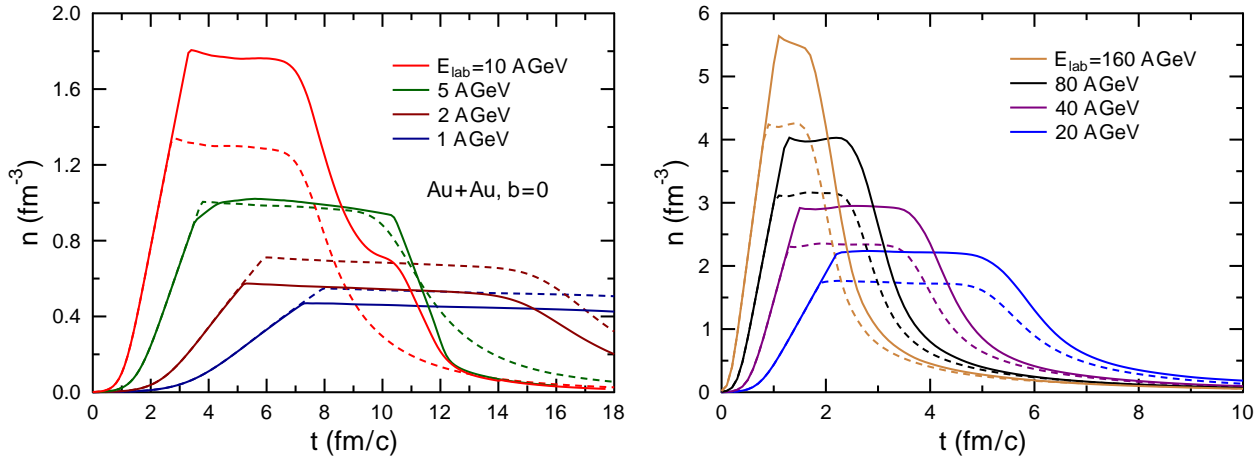}}
\caption{(Color online)
Same as Fig.~\ref{fig4}, but for net baryon density.
}
        \label{fig5}
        \end{figure*}
\begin{figure*}[htb!]
        \centerline{\includegraphics[width=0.65\textwidth]{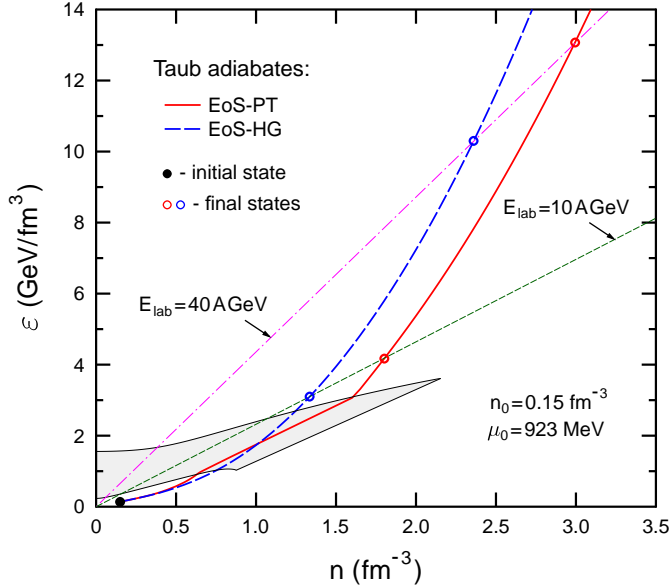}}
\caption{(Color online)
Taub adiabates and parameters of shock waves for collisions
of two slabs of normal nuclear matter for $\ela=10$ and 40 AGeV.
Thin straight lines represent stopping conditions~(\ref{stop}) for these values of
bombarding energy.
Thick solid and dashed lines are the Taub adiabates (\ref{taub}) for \mbox{EoS--PT}
and \mbox{EoS--HG}, respectively. The full dot shows the initial state $n=n_0, T=0$. Open dots
represent parameters of compressed zone behind the shock fronts. Shading
shows the MP region of deconfinement phase transition.
}
        \label{fig6}
        \end{figure*}

Below we discuss thermodynamic characteristics of matter created in Au+Au collisions
at various bombarding energies.
Our goal is to find differences between the results obtained with
EoS--HG and EoS--PT. Let us first consider purely central collisions with $b=0$.
Figures~\ref{fig4}--\ref{fig9} show the time evolution of the rest
frame energy and baryon densities averaged over the central box defined in Sec.~\ref{snums}.
One can see that after a rapid initial growth the baryon and energy densities
remain practically constant during certain time intervals (depending on the bombarding energy $\ela$)
and only later they start to decrease.
The plateau is explained by running shock waves created at an early stage of the reaction.
Within the considered one-fluid approach the entropy of matter is generated
via the shock--like compression mechanism (see the next section)
which converts the c.m. kinetic energy of
projectile and target nucleons into internal energy of stopped matter.
The plateaus in Figs.~\ref{fig4}--\ref{fig5} begin at
the time moments when shock waves reach dense central parts of colliding nuclei.
Initially these shocks appear at the central point $\bm{r}=0$ in the c.m. frame and then propagate
outwards through the colliding nuclei.
Right-end points of the plateaus correspond to the moments when rarefaction waves, propagating
from rear sides of nuclei, reach the central point. One can see that at \mbox{$\ela\gtrsim 5$ AGeV},
when matter enters MP states, the calculation with the EoS--PT predict larger values of
$\varepsilon$ and $n$ as compared with the EoS--HG.

\subsection{Simplified picture of matter evolution in a central box}\label{dyntr}
        \begin{figure*}[htb!]
        \vspace*{-1cm}
        \centerline{\includegraphics[width=0.55\textwidth]{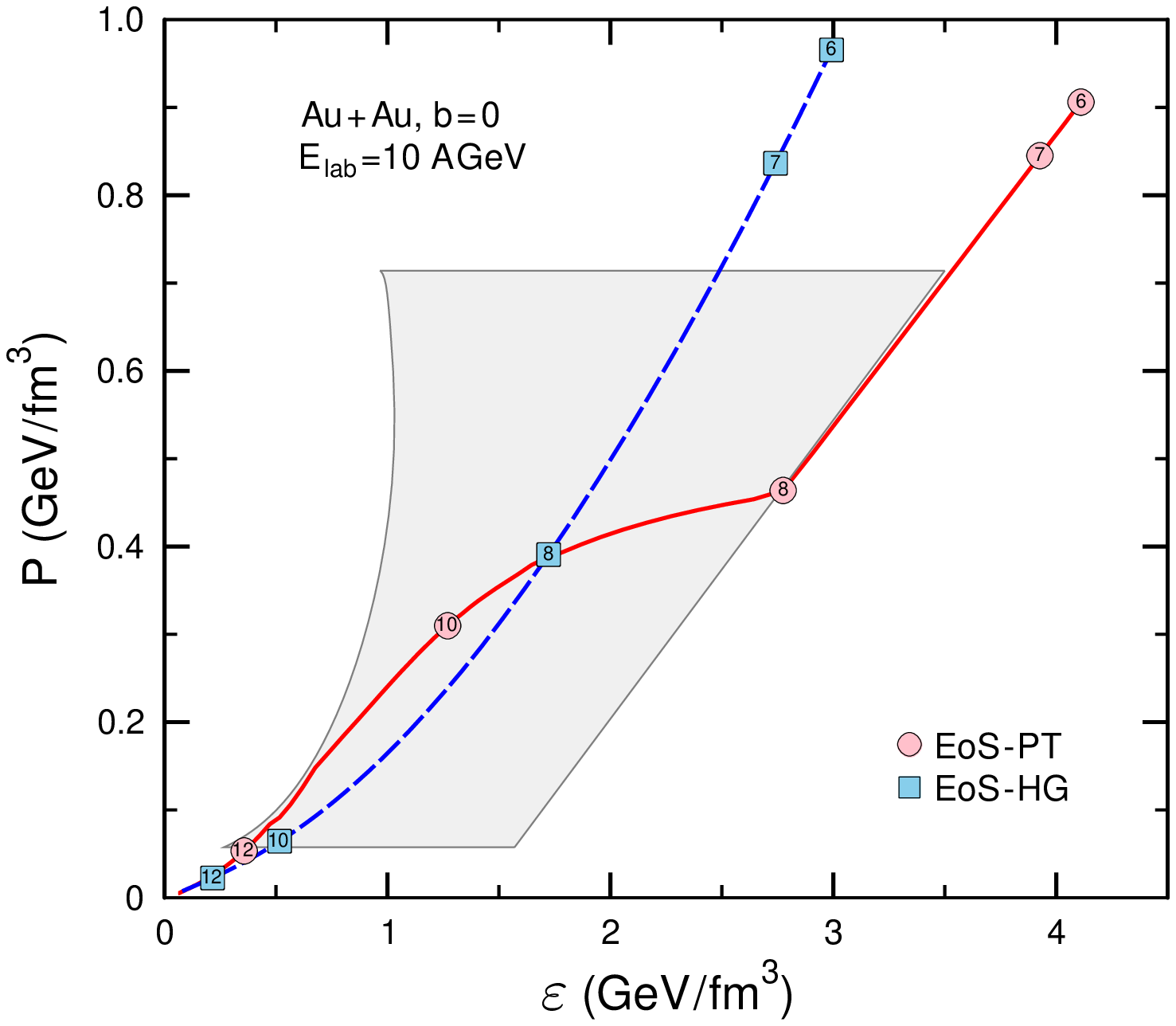}}
        \caption{(Color online)
        Pressure as a function of energy density in a central box for central 10 AGeV Au+Au collisions.
        Dashed and solid lines give the results for EoS--HG and EoS--PT, respectively.
        Numbers in squares and circles
        denote the values of c.m.~time (shown are only states
        at \mbox{$t>6$ fm/$c$}). Shading represents the MP region.
        }
        \label{fig7}
        \end{figure*}
To estimate parameters of states with maximal compression,
let us consider a collision of two slabs of cold nuclear matter  (\mbox{$n=n_0, T=0$})
moving initially with velocities $\pm\hsp v_0$ along the $z$--axis.
After the first contact two shock fronts start to propagate with constant velocities
in the positive and negative $z$--directions.
From the continuity of $T^{\hsp 0z}, T^{\hsp zz}, N^z$ in the shock front rest frame
one gets~\cite{Lan87} the relation (so-called ''Taub adiabate'') connecting the baryon~($n$)
and energy ($\varepsilon$) densities of fluid behind the shock front:
\bel{taub}
\varepsilon_0\hsp (P+\varepsilon_0)\hsp n^2=\varepsilon\hsp (P+\varepsilon)\hsp n^2_0\hsp,
\ee
where $\varepsilon_0=\mu_0\hsp n_0$ is the initial rest--frame energy density of each
slab\,\footnote
{
The explicit form of \re{taub} is obtained assuming that the initial pressure
$P_0=P(n_0,\varepsilon_0)=0$.
}.

In addition one can write down the equation expressing conservation
of energy per baryon at the shock front. In the c.m. frame this equation
looks as the stopping condition
\bel{stop}
\varepsilon/n=\gamma_0\hsp\varepsilon_0/n_0\hsp,
\ee
where $\gamma_0=(1+\ela/2m_N)^{1/2}$ is the initial Lorentz factor.
Graphic solutions of Eqs.~\mbox{(\ref{taub})--(\ref{stop})} for two EoS\hspace*{0.5pt}s used in this paper are
shown in Fig.~\ref{fig6}. One can see that at $\ela\gtrsim 10$ AGeV\ calculations with the
EoS--PT predict noticeably higher maximal values of $\varepsilon$ and $n$ as compared with
the EoS--HG. This in turn leads to higher collective flows of matter
in hydro simulations with the phase transition.
Comparison with parameters of maximal compression
in central Au+Au collisions (see Figs.~\ref{fig4}--\ref{fig5}) shows a good agreement
between the Taub adiabate predictions and fluid dynamical calculations.
Our shock wave calculations show that phase transition effects and, in particular,
excitation of MP states
may be expected only at sufficiently high bombarding energies, $\ela\gtrsim 3$ AGeV\,\footnote
{
According to Figs.~\ref{fig4}--\ref{fig5}, there are some differences between
the EoS--PT and EoS--HG calculations at energies \mbox{$\ela=1-2$ AGeV}. This is a consequence
of excluded volume effects disregarded in the EoS--HG.
}.

Decreasing parts of curves in Figs.~\ref{fig4}--\ref{fig5} correspond to
the expansion stage of system evolution in the central box. It can be shown that
this expansion proceeds quasi--adiabatically, i.e. with nearly constant entropy per baryon $\sigma$
(small deviations are caused by numerical viscosity and by averaging over cells in
the central box).
Figure~\ref{fig7} shows time evolution of thermodynamic parameters in the central box
for the case of central Au+Au collisions at \mbox{$\ela=10$ AGeV}. Only late stages
of the reaction in the $\varepsilon-P$ plane are considered. Comparison with trajectories
in Fig.~\ref{fig2} shows that in the considered case the fluid expansion in the central box
proceeds along the adiabate with $\sigma\sim 10$ (for both EoS\hspace*{0.5pt}s).
From Figs.~\ref{fig4}--\ref{fig5},~\ref{fig7} one can see that in the case of
the EoS--PT the system expands with some delay as compared with the EoS--HG, i.e.
the same values of $\varepsilon$ and $n$ are achieved at later times. However,
due to differences in pressure, final hadronic states predicted for the EoS--PT and EoS--HG have,
generally speaking, different collective velocities.

\subsection{Comparison of one and three--fluid models}

        \begin{figure*}[htb!]
        \vspace*{-1.5cm}
        \centerline{\includegraphics[width=0.6\textwidth]{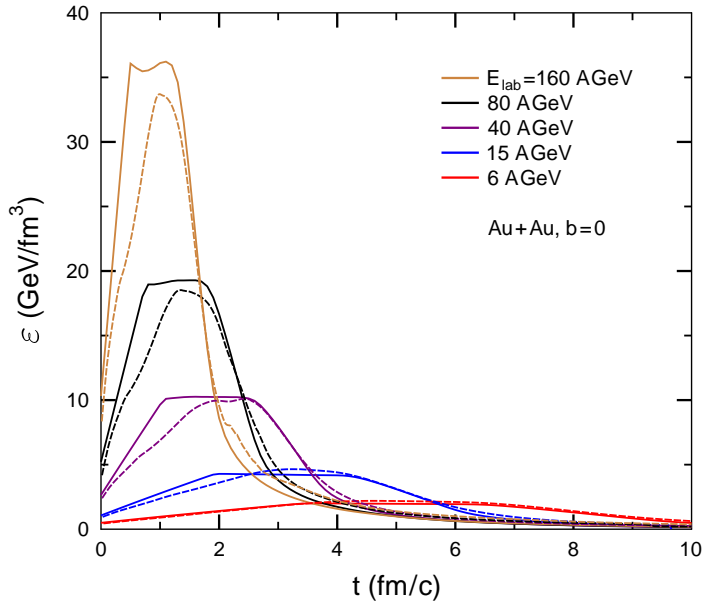}}
        \caption{(Color online)
        Time evolution of energy density in the central box of central Au+Au collisions
        with different~$\ela$. Solid lines shows our results in the case of EoS--HG and sharp
        density distributions of initial nuclei. Dashed lines represent three--fluid
        calculations~\cite{Iva09b}.
        }
        \label{fig8}
        \end{figure*}
It should be stressed that the results presented above were obtained under the assumption
of full mutual stopping of projectile and target matter.
Presumably, this approximation should be less justified with raising bombarding energies
(see a more detailed discussion in~\cite{Mis91}). To estimate the importance of
projectile--target transparency effects,
we have compared our calculations with the results of the three--fluid model of Ref.~\cite{Iva06}.
The standard version of this model applies a purely hadronic EoS.
In addition, sharp density profiles of initial nuclei are assumed. To make the comparison possible,
we performed our calculations with the EoS--HG choosing a small diffuseness parameter $a=0.01$ fm
(see~\re{dinit}). The results of this comparison are presented in Fig.~\ref{fig8}.
One can see that in the case of central Au+Au collisions the differences between the one-- and
three--fluid calculations are rather small at \mbox{$\ela\lesssim 30$ AGeV}.
At larger bombarding energies maximal values of $\varepsilon$ predicted by one--fluid calculations
are about 10\% larger than in three--fluid model. In addition, the latter predicts a noticeably
slower increase of energy density at the initial stage of the collision.
From these results we conclude that transparency effects are not so
strong at least in central Au+Au collisions at FAIR energies and below.

\subsection{Space--time picture of central Au+Au collision}
        \begin{figure*}[htb!]
        \hspace*{-0.5cm}\includegraphics[width=0.65\textwidth]{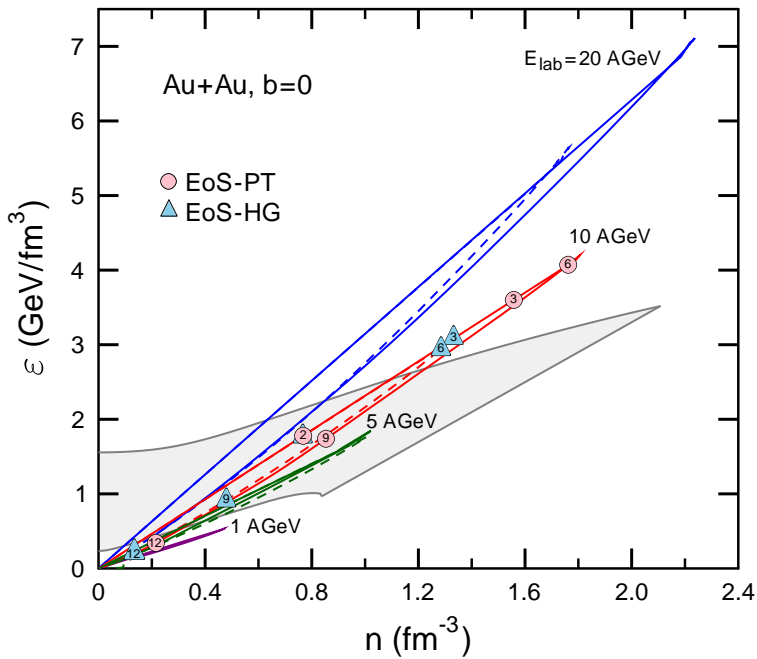}
        \caption{(Color online)
        Time evolution of matter in central Au+Au collision at different $\ela$.
        Shown are values of energy and baryon densities averaged over the central box.
        Dashed and solid lines correspond to EoS--HG and EoS--PT, respectively. Numbers
        in triangles and circles give the c.m. time in fm/$c$.
        Shading shows MP region of deconfinement phase transition.
        }
        \label{fig9}
        \end{figure*}
        \begin{figure*}[htb!]
        \centerline{\includegraphics[width=0.85\textwidth]{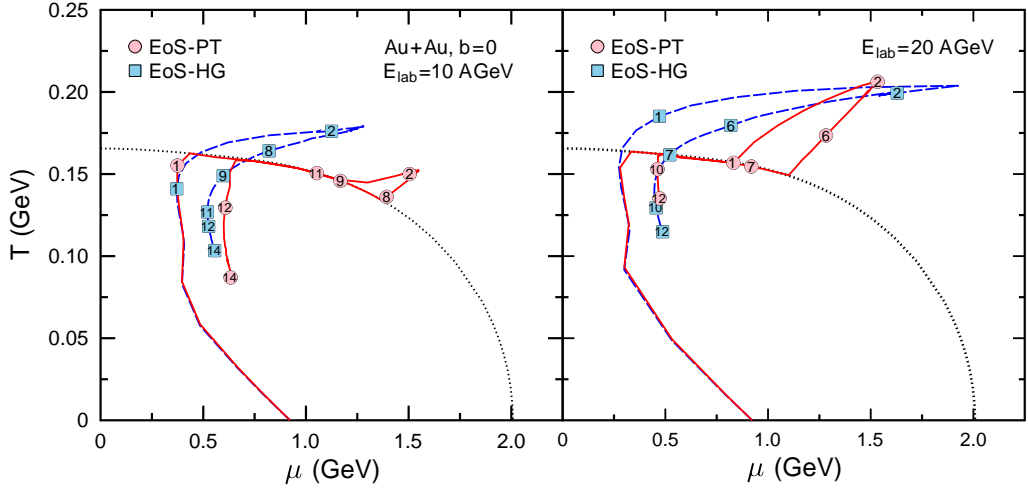}}
        \caption{(Color online)
        Time evolution of matter (central cell) in the \mbox{$\mu-T$} plane.
        Left and right panels correspond to central Au+Au collisions
        at $\ela=10$ and 20 AGeV, respectively.
        Dashed (solid) lines show the results for EoS--HG (EoS--PT). Numbers
        in squares and circles give time values in fm/$c$.
        }
        \label{fig10}
        \end{figure*}
       \begin{figure*}[htb!]
        \centerline{\includegraphics[width=0.9\textwidth]{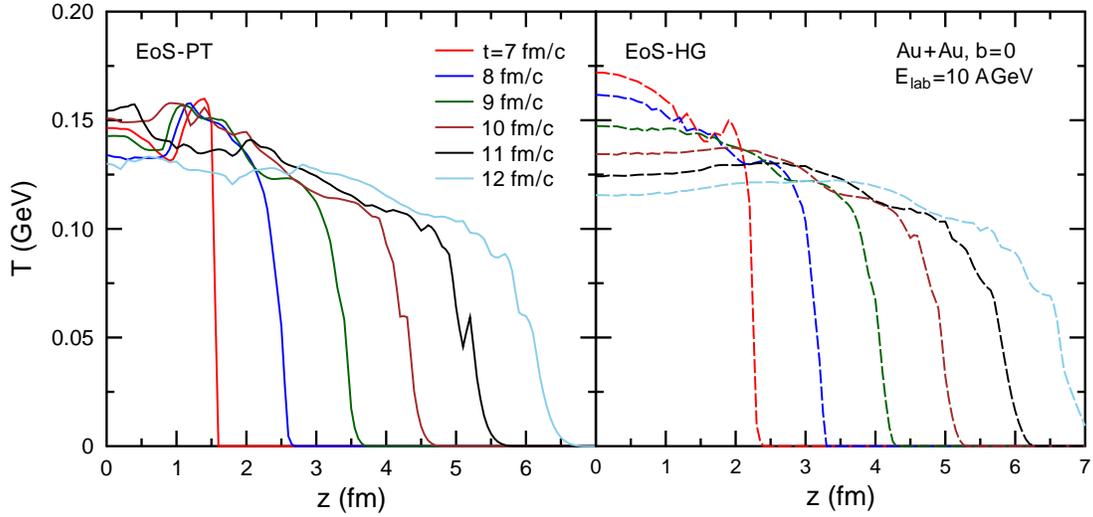}}
        \caption{(Color online)
        The temperature profiles along the beam axis at different time moments
        in a central 10 AGeV Au+Au collision. The left (right) panel corresponds to EoS--PT (EOS--HG).
        The results for $z>0$ are shown only.
        }
        \label{fig11}
        \end{figure*}
       \begin{figure*}[htb!]
        \centerline{\includegraphics[width=0.9\textwidth]{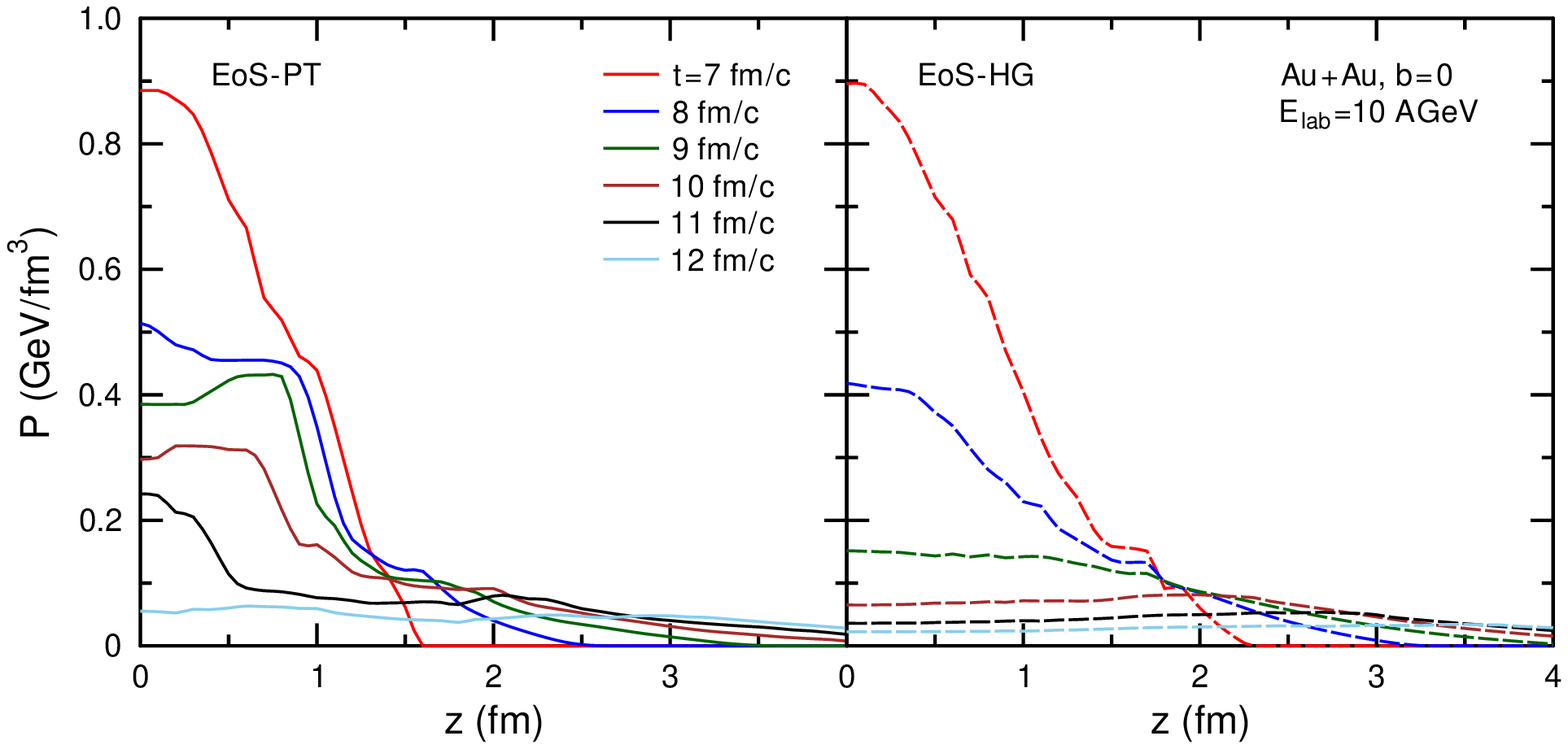}}
        \caption{(Color online)
        Same as Fig.~\ref{fig11} but for the pressure profiles along the beam axis.
        }
        \label{fig11a}
        \end{figure*}
Figures~\ref{fig9}--\ref{fig10} represent dynamical
trajectories of matter (the central box) produced in central Au+Au collisions at
different bombarding energies. Shown are results for $n-\varepsilon$ and $\mu-T$ planes,
respectively. As compared with the EoS--HG, the calculations with the phase transition
predict longer life-times of states with maximal compression and additional time delays
of expansion stages.
Especially large differences between two EoS\hspace*{0.5pt}s are visible at intermediate times in
the $\mu-T$ plane.
The dotted line in Fig.~\ref{fig10} shows the MP region in this plane~\footnote
{
As discussed in~\cite{Ton04}, due to effects of strangeness neutrality,
the region of MP states is not a line, as in the case of one conserved charge,
but a strip. However, at $v_e\lesssim 1$ fm$^3$ the latter
is rather narrow~\cite{Sat09} and practically not distinguishable from the dotted line in
Fig.~\ref{fig10}.
}.
The calculations with EoS--PT predict a zigzag-like behavior of the trajectories in the $\mu-T$ plane
with slightly raising temperature in the MP as a function of time. On the other hand, the trajectories
do not bend at the MP boundaries in the $n-\varepsilon$ plane. According to
Figs.~\ref{fig9}--\ref{fig10}, the largest life times of the MP in central Au+Au collisions,
about~4 fm/$c$, are reached at $\ela\sim 10$ AGeV. At lower energies the MP region is traversed only partly,
while at higher $\ela$ the produced matter expands too fast.
        \begin{figure*}[thb!]
        \centerline{\includegraphics[width=0.55\textwidth]{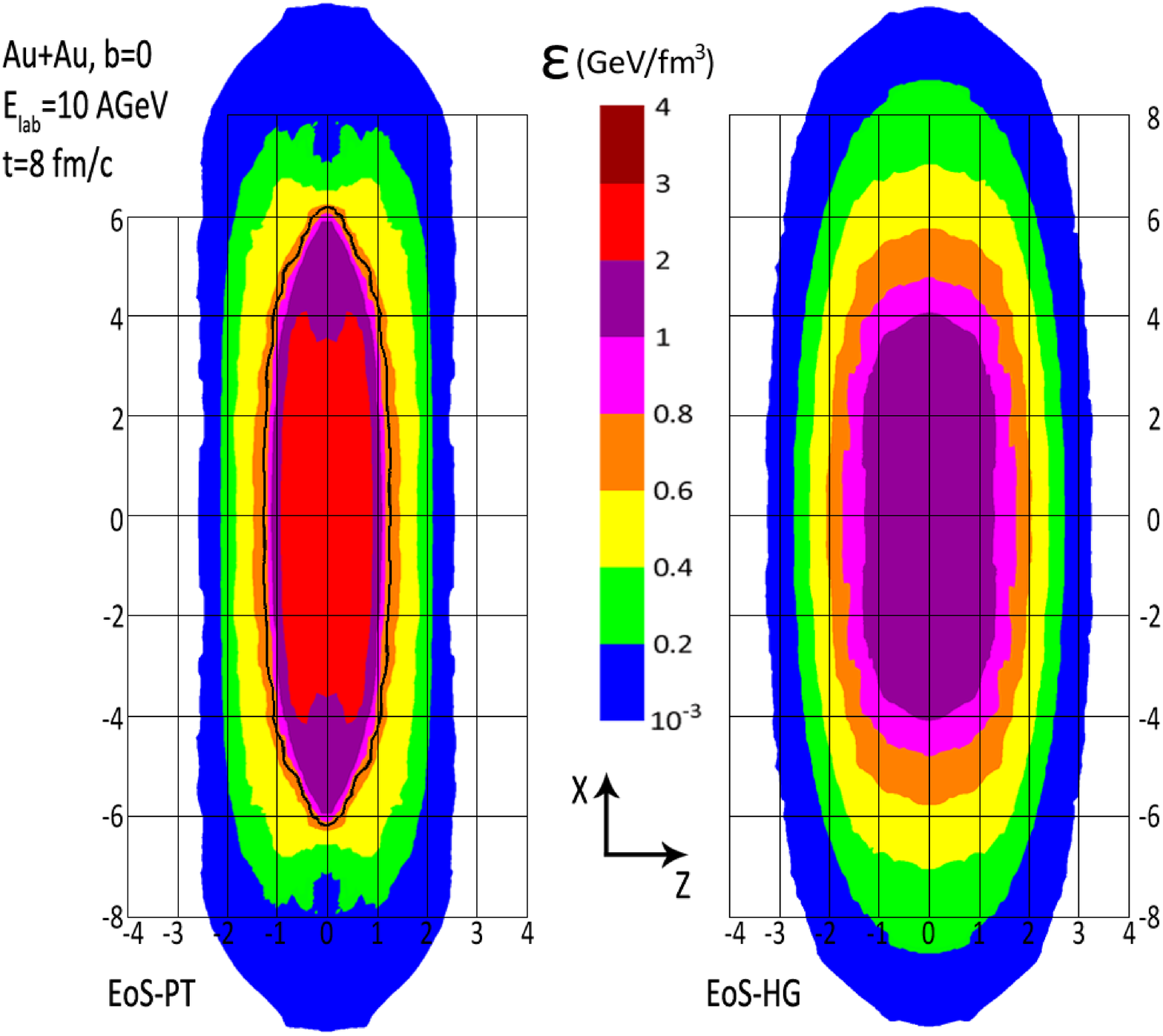}}
        \caption{(Color online)
        Contours of equal energy densities (in GeV/fm$^3$) in the reaction ($y=0$)
        plane at $t=8$ fm/$c$ for central 10 AGeV Au+Au collision.
        The left (right) panel corresponds to EoS--PT (EoS--HG).
        Numbers near the vertical and horizontal lines show the $x$ and $z$
        coordinates in fm. Thick line in the left panel shows the boundary of MP.
        }
        \label{fig12}
        \end{figure*}

        \begin{figure*}[b!]
        \vspace*{3mm}
        \centerline{\includegraphics[width=0.65\textwidth]{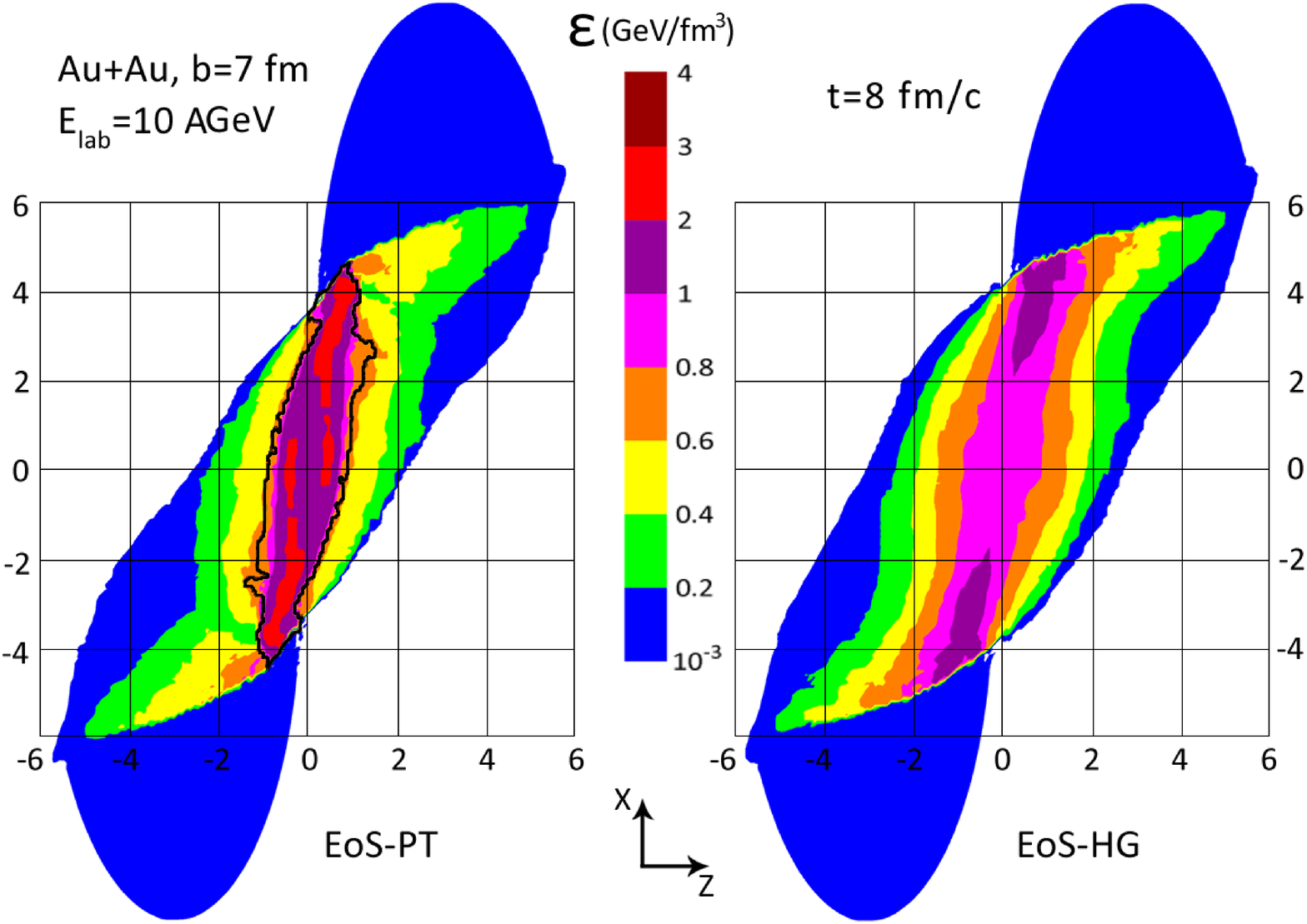}}
        \caption{(Color online)
        Contours of equal energy densities in the reaction ($y=0$)
        plane at $t=8$~fm/$c$ for 10 AGeV Au+Au collision with $b=7$ fm.
        The left (right) panel corresponds to EoS--PT (EoS--HG).
        Numbers near the vertical and horizontal lines show the $x$ and $z$
        coordinates in fm. Thick line in the left panel shows the boundary of MP.
        }
        \label{fig13}
        \end{figure*}
Figures~\ref{fig11}--\ref{fig11a} shows the temperature and pressure profiles as functions of the longitudinal
coordinate along the beam axis (\mbox{$x=y=0$}) for central Au+Au collision
at \mbox{$\ela=10$ AGeV}. It is interesting that in the case of EoS--PT the temperature at small
$|z|$ changes non-monotonically in the interval \mbox{$t=8-12$ fm/$c$}\hsp.  This follows from a specific
behavior of the critical temperature as a function of $\mu$. Indeed, as one can see from
Fig.~\ref{fig10}, the temperature of MP in a central slice increases noticeably during the expansion.
Such a behavior is explained by the release of the latent heat in hadronizing process.
According to Fig.~\ref{fig11a}, the deconfinement phase transition leads to slower drop of pressure gradients
in expanding matter along the beam direction. These gradients are especially large at the MP boundary.
This results in strong acceleration of fluid cells during the hadronization processes.
As will be shown below, such acceleration leads to additional broadening of baryon rapidity distribution
as compared to a purely hadronic scenario (see Sec. V).

Comparison of spatial distributions of energy density for two EoS\hspace*{0.5pt}s is made
in Fig.~\ref{fig12}. It reprents the contour plots of $\varepsilon$ in the reaction plane of a 10 AGeV central
Au+Au collision at $t=8$ fm/$c$\hsp. The calculation with the EoS--PT shows that at this time the central
zone is still in the MP. In this case maximal values of energy density are noticeably
higher as compared with the EoS--HG.
        \begin{figure*}[htb!]
        \centerline{\includegraphics[width=0.65\textwidth]{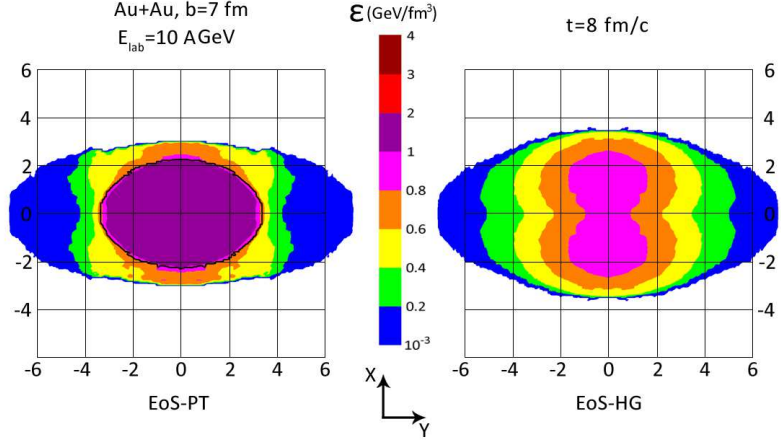}}
        \caption{(Color online)
        Same as Fig.~\ref{fig13}, but for the transverse plane $z=0$.
        }
        \label{fig14}
        \end{figure*}

\section{Modeling non-central collisions}

\subsection{Energy density and collective velocities in Au+Au collisions at $\ela = 10$ AGeV}

       \begin{figure*}[htb!]
        \vspace*{5mm}
        \centerline{\includegraphics[width=0.75\textwidth]{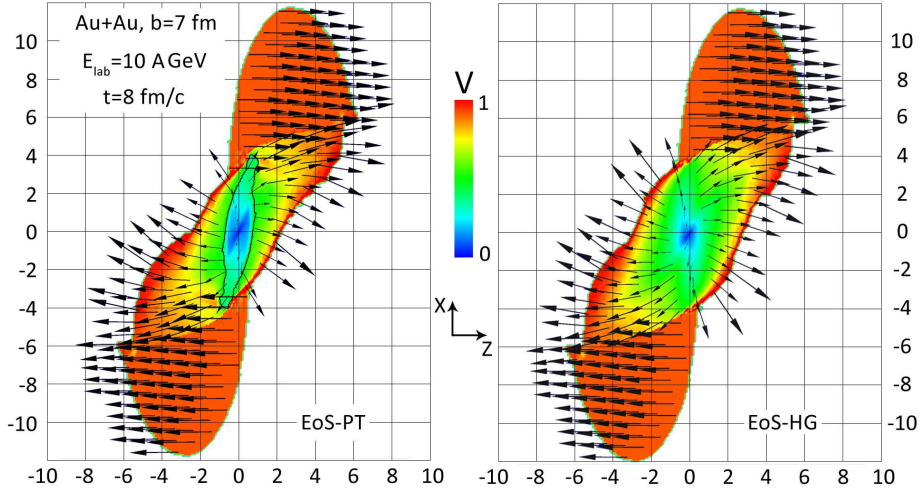}}
        \caption{(Color online)
        Collective velocities of matter in the reaction plane
        for midcentral 10~AGeV Au+Au collision ($t=8$ fm/$c$, \mbox{$b=7$ fm}).
        Left and right panels correspond to EoS--PT and EoS--HG respectively.
        Lengths and directions of arrows represent 3--velocity vectors. Numbers show the $z$ and $x$
        coordinates in fm. Thick line in the left panel shows the boundary of MP.
        Vertical color strips characterize modules of \mbox{3--velocity} in units of $c$.
        }
        \label{fig15}
        \end{figure*}
Let us consider now non-central collisions.
As we shall see below, especially interesting is the region of  bombarding energies $\sim 10$ AGeV.
At such energies we expect an enhanced sensitivity of collective flow observables
and particle spectra to the phase transition.
In this section we present results for 10 AGeV Au+Au collisions with the impact parameter~\mbox{$b=7$ fm}.

\begin{figure*}[htb!]
        \centerline{\includegraphics[width=0.6\textwidth]{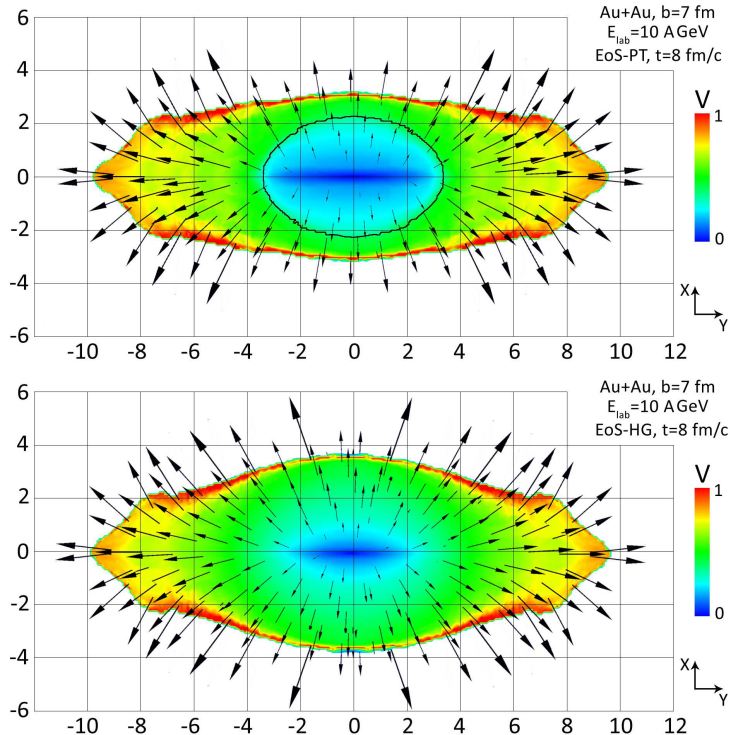}}
        \caption{(Color online)
        Collective velocities of matter in the transverse plane
        for 10 AGeV Au+Au collision ($t=8$ fm/$c$, \mbox{$b=7$ fm}).
        Upper and lower panels correspond to EoS--PT and EoS--HG respectively.
        Numbers show the $x$ and $y$ coordinates in fm. Thick line in the upper panel shows
        the boundary of MP. Vertical color strips characterize modules of \mbox{3--velocity}
        in units of $c$.
        }
        \label{fig16}
        \end{figure*}

Figures~\ref{fig13}--\ref{fig16} show contour plots of the energy density
and 3-velocity fields in the reaction~($z-x$) and transverse ($y-x$) planes
at $t=8$ fm/$c$. One can see that in calculation with the EoS--PT a significant
part of the system at this stage is still in the MP.

Figures~\ref{fig15}--\ref{fig16} represent the collective velocity fields for the same reaction.
Blue colored central regions in these figures correspond to spatial domains with relatively low
absolute values of c.m. velocity $v\lesssim 0.2\,c$. At the considered time moment, such a region
occupies much larger fraction of space in the calculation with the deconfinement phase transition.
According to \mbox{Figs.~\ref{fig15}--\ref{fig16}}, the MP matter is practically
at rest in the c.m. frame. On the other hand, pronounced collective flows of hadronic
matter are formed in transverse directions. Especially large transverse velocity
components appear in the region of geometrical overlap of initial nuclei, at outer edges
of the HP. Qualitative structure of the velocity fields may be understood if
one takes into account that fluid elements are accelerated mainly along the directions
of largest gradients of energy density (see Sec.~\ref{EoS}).
In particular, this leads to the effect of ''antiflow''
i.e. to appearance of fluid elements with opposite signs of $v_x$ and~$v_z$
in the reaction plane (see Fig.~\ref{fig15}).
        \begin{figure*}[htb!]
        \centerline{\includegraphics[width=0.85\textwidth]{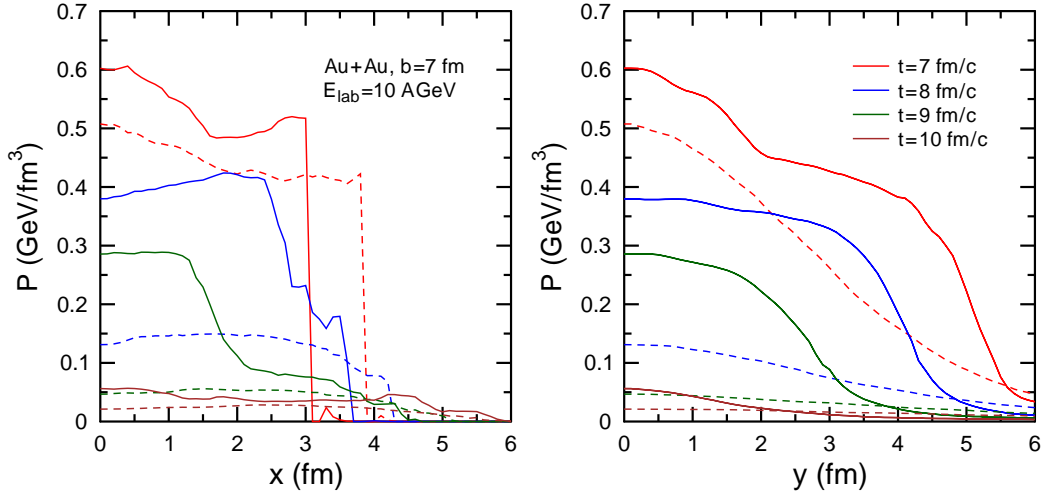}}
        \caption{(Color online)
The pressure profiles in 10 AGeV Au+Au collision ($b=7$ fm).
Left and right panels correspond, respectively, to the $x$ and $y$ axis in the transverse
plane $z=0$. Red and blue curves give the results for $t=8$ and $9$ fm/$c$.
Solid and dashed lines are calculated with EoS--PT and EoS--HG, respectively.
        }
        \label{fig17}
        \end{figure*}

According to Fig.~\ref{fig14}, at~\mbox{$\ela\simeq 10$ AGeV} the EoS--PT leads to
an essentially different spatial symmetry of the energy density distribution
in the transverse plane as compared with the purely hadro\-nic~EoS.
On the other hand, the calculation with the EoS--PT predicts
much larger pressure gradients and accelerations of matter in the
$x$ and $y$ directions. In Fig.~\ref{fig17} we compare the pressure profiles
for both EoS\hspace*{0.5pt}s at different times. One can see, that in the case of the
EoS--PT, especially strong pressure gradients at $t\sim 8$ fm/$c$ are formed in
the $x$ direction, at the boundary between the hadronic and mixed phases.
This in turn leads to an enhanced momentum anisotropy in the case of EoS--PT
as compared with the EoS--HG (see next section)~\footnote
{
Essentially the same effect is responsible for the broadening of baryon rapidity
distribution in central collisions as discussed in Sec. III\hsp D and V.
}.

\subsection{The momentum anisotropy at different bombarding energies}\label{moman}

\begin{figure*}[htb!]
        \centerline{\includegraphics[width=\textwidth]{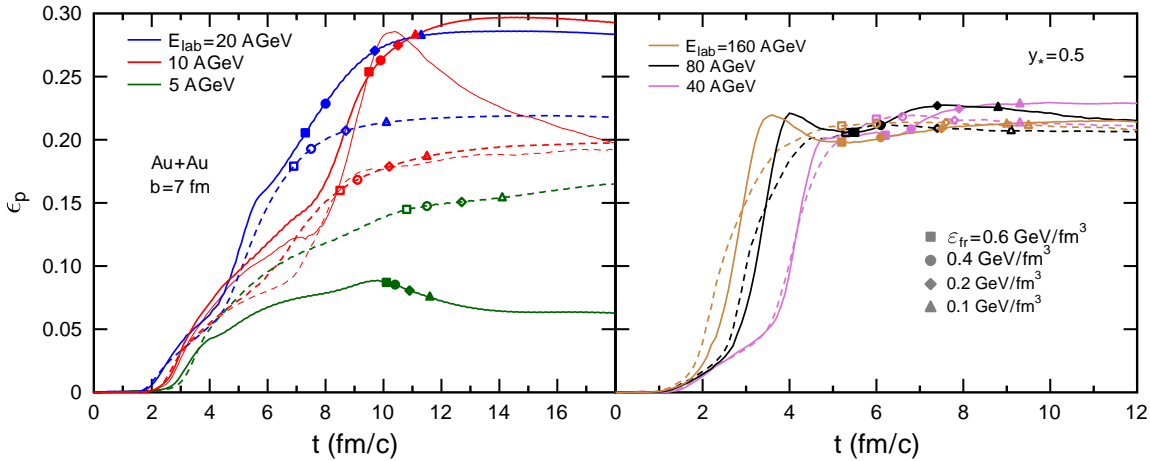}}
        \caption{(Color online)
        Time dependence of the momentum anisotropy
        in Au+Au collisions at different bombarding energies ($b=7$ fm).
        The solid (dashed) lines correspond to EoS--PT (\mbox{EoS--HG}). Dots
        marks time moments when the energy density in central box
        becomes lower certain values~$\varepsilon_{\rm{fr}}$. Thin lines
        in the left panel show the results of calculations for $\ela=10$ AGeV
        with simplified definition of $\epsilon_p$ (see text).
        }
        \label{fig18}
        \end{figure*}

Here we present the results concerning the momentum anisotropy $\epsilon_p$
in (semi)peripheral Au+Au collisions. This quantity has been defined in \re{anism}.
Figure~\ref{fig18} represents the time evolution
of $\epsilon_p$ at different $\ela$\,. Thick lines are calculated using the generalized
definition of $\epsilon_p$ with $y_*=0.5$ (see footnote~[7]). By thin lines (for $\ela=10$ AGeV)
we also present calculations in the limit \mbox{$y_*\to 0$}. One can see that in the case with phase transition
the simplified formulae (\ref{anism}) leads to unphysical maximum of $\epsilon_p (t)$ at intermediate
times. According to Fig.~\ref{fig18}, the momentum anisotropy is rather sensitive to
EoS in the bombarding energy range \mbox{$\ela\lesssim 20$ AGeV}. The asymptotic values of $\epsilon_p$
at $\ela\gtrsim 10$ AGeV are larger in calculations with the deconfinement phase transition.
We think that such ''counterintuitive'' result\,\footnote
{
A similar conclusion about the increase of elliptic flow due to
phase transition effects at FAIR energies has been made recently in Ref.~\cite{Pet10}\label{footep}
}
may be explained by the difference in pressure gradients in the transverse plane.
As discussed in the end of preceding section, they are much larger at intermediate times
if the MP is created.
        \begin{figure*}[htb!]
        \centerline{\includegraphics[width=0.6\textwidth]{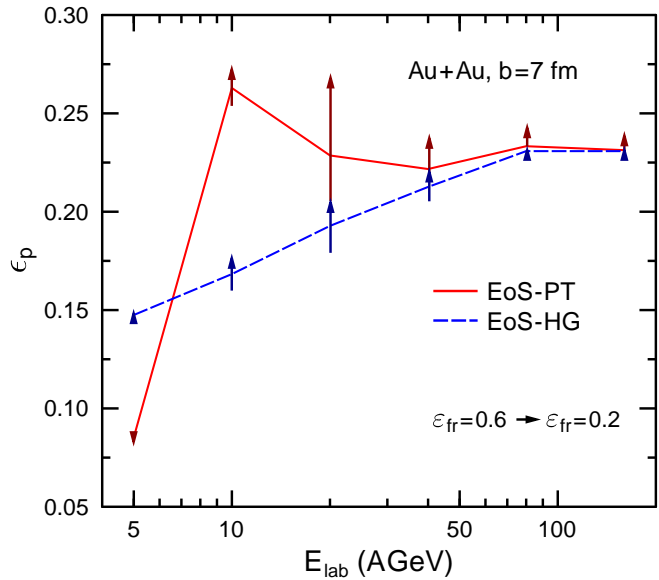}}
       \vspace*{-5mm}
        \caption{(Color online)
        Excitation function of momentum anisotropy in Au+Au collisions with
        \mbox{$b=7$ fm}.  The solid (dashed) lines correspond to EoS--PT (EoS--HG).
        Arrows show possible shifts of $\epsilon_p$--values for different choices
        of freeze-out energy density $\varepsilon_{\rm fr}$ between 0.2 and 0.6 GeV/fm$^3$.
        }
        \label{fig19}
        \end{figure*}

As mentioned in Sec.~\ref{specf} the freeze--out effects may significantly change
the observed values of transverse flows in heavy--ion collisions. To estimate
a possible spread of results we determine the time moment $t_{\hsp\rm fr}$ when the energy density
in a central box becomes smaller a~certain freeze-out value $\varepsilon_{\rm fr}$.
Applying such freeze-out condition, in Fig.~\ref{fig18}
we mark by points the values of $\epsilon_p$ corresponding to different values
of $\varepsilon_{\rm fr}$. Figure~\ref{fig19} shows excitation functions of the momentum
anisotropy for two EoS\hspace*{0.5pt}s. The lines connect the values of $\epsilon_p$ taken at
$t=t_{\hsp\rm fr}$ where $t_{\hsp\rm fr}$ corresponds to \mbox{$\varepsilon_{\rm fr}=0.4$ GeV/fm$^3$}.
At each $\ela$ the ends of arrows in Fig.~\ref{fig19} show the values of~$\epsilon_p$ for
\mbox{$\varepsilon_{\rm fr}=0.2$} and $0.6$ GeV/fm$^3$. In the case of EoS--PT our model
predicts a non-monotonic dependence $\epsilon_p\hsp (\ela)$ with maximum at \mbox{$\ela\simeq 10$ AGeV}.

As indicated above, the momentum anisotropy is approximately proportional to the pion elliptic flow at
midrapidity. It is interesting that existing experimental data on the proton and pion elliptic
flows in Au+Au and Pb+Pb collisions (see their compilation in Ref.~\cite{Iva09a}) do not exclude presence
of a local maximum of $v_2$ at high AGS energies. However, the data at AGS and SPS energies were obtained
by using different detectors, centrality cuts and methods of~$v_2$ determination. It would be highly desirable
to perform more detailed measurements of the elliptic flow excitation function in the bombarding energy range \mbox{5--50~AGeV}.
Hopefully, low energy RHIC as well as future NICA and FAIR experiments will help in obtaining such data.

\section{Hadronic spectra}

        \begin{figure*}[htb!]
        \centerline{\includegraphics[width=0.9\textwidth]{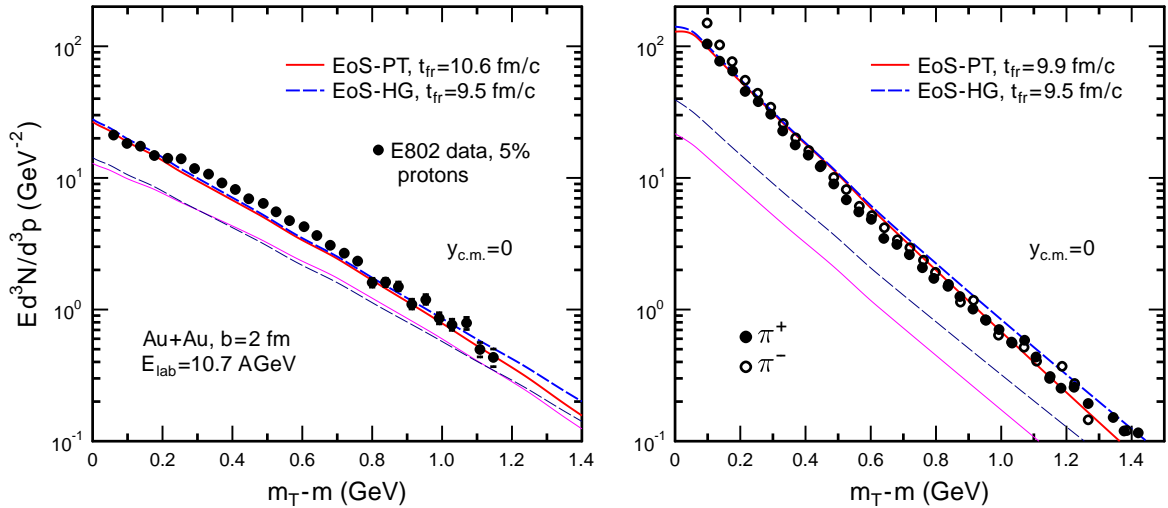}}
        \caption{(Color online)
        Transverse kinetic energy distributions of protons (left panel) and negative pions (right panel)
        in 10.7 AGeV Au+Au collision ($b=2$ fm). Thick solid and dashed lines are calculated
        with EOS--PT and EOS-HG, respectively. Thin lines show the distributions of thermal hadrons,
        i.e. without contributions from decays of resonances. Experimental data are taken from~\cite{Ahl98}.
        }
        \label{fig20}
        \end{figure*}
        \begin{figure*}[htb!]
        \vspace*{-5mm}
          \centerline{\includegraphics[width=0.65\textwidth]{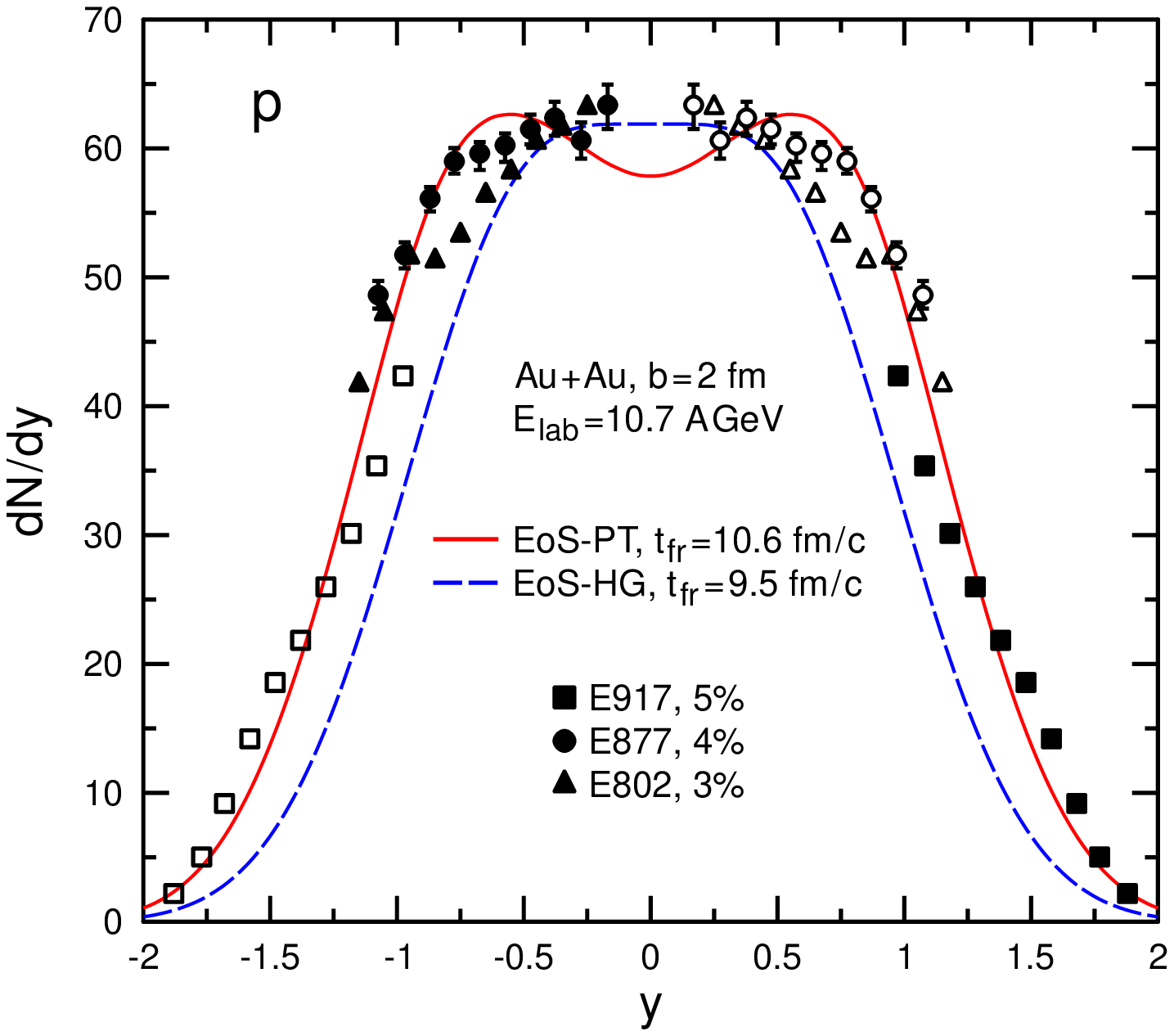}}
        \vspace*{-5mm}
        \caption{(Color online)
        Rapidity distributions of protons in 10.7 AGeV Au+Au collision ($b=2$~fm).
        Solid and dashed lines are calculated with EOS--PT and EOS-HG, respectively.
        Full dots are experimental data~\cite{Ahl98,Bac01,Bar00}, circles are obtained by
        reflection with respect to midrapidity.}
        \label{fig21}
        \end{figure*}
In this section we present the model results for proton and pion momentum distributions
in central Au+Au collisions at \mbox{$\ela\simeq 10$ A GeV}. These distributions are
calculated by using the Cooper-Frye formula~(\ref{spec}) with the
isochronous freeze-out hypersurface. The results for the EoS--PT include the excluded volume
corrections (see Sect.~\ref{specf}). The proton and $\pi^-$  distributions
are obtained from nucleon and pion spectra by introducing the scaling factors~1/2 and
1/3, respectively. The parameter of freeze--out time $t_{\rm fr}$ is chosen
to achieve the best fit of experimental data~\footnote
{
To reduce influence of isospin effects, we fitted our distributions of
''negative'' pions to one half of observed charged pion distributions.
}.
Our calculations with the EoS--PT show that the best choice
corresponds roughly to the end of hadronization stage.
In this case the thermodynamic parameters of central cells become
close to the boundary between the MP and HP (see Figs.~\ref{fig9}--\ref{fig10}).

According to our analysis, significant fractions of proton and pion yields in the considered
reaction come from decays of resonances. At $\ela\sim 10$ AGeV,  relative contributions of free
(''thermal'') pions and protons at midrapidity are about
25 and 60 percents of total yields of pions and protons, respectively. Especially important are
contributions of isobar decays. This is explained by large values of baryon chemical potential
(\mbox{$\mu\sim 0.6-0.7$ GeV} at freeze-out) expected in nuclear collisions at AGS and FAIR energies.
Having in mind that measured yields include feeding from weak decays
of hyperons and $K^{\hsp 0}_S$ mesons, we take into account the corresponding
contributions in our calculations.

        \begin{figure*}[htb!]
          \centerline{\includegraphics[width=0.65\textwidth]{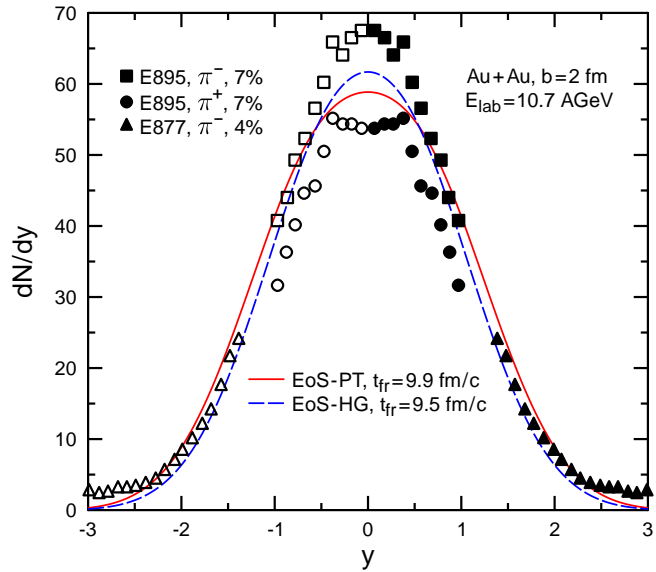}}
        \vspace*{-5mm}
        \caption{(Color online)
        Same as Fig.~\ref{fig19} but for rapidity distributions of $\pi^-$ mesons. Full symbols are
        experimental data for charged pions in central Au+Au collisions
        at $\ela=8$~\cite{Kla03} and $9.9$~\cite{Bar00}~AGeV. Open symbols are obtained by
        reflection.}
        \label{fig22}
        \end{figure*}
According to Fig.~\ref{fig20}, the proton and pion $p_T$  distributions in central Au+Au collisions at
\mbox{$\ela\sim 10$ AGeV} are not so sensitive to EoS. In this figure we explicitly show the contributions
of ''directly'' produced (thermal) proton and pions. Indeed, one can see that resonance decays
give large contributions at all $p_T$, especially for pions.

The proton rapidity distribution for the same reaction is presented in
Fig.~\ref{fig21}. As compared to the EoS--HG, the calculation with the phase transition predicts
a noticeably broader rapidity distribution.
In this case the agreement with experimental data outside the central rapidity region is better
than in the purely hadronic scenario.
The physical meaning for this broadening is related to larger pressure gradients in the
longitudinal direction, as demonstrated in Fig.~\ref{fig11a}.
Our analysis shows that the shape of the proton rapidity
distribution at small c.m. rapidities is rather sensitive to the excluded volume parameter.
It is interesting that single-- and three--fluid calculations~\cite{Bra97}
of central 11 AGeV Au+Au collisions also predict appearance of a deep in the proton rapidity distribution at
midrapidity for a EoS with the deconfinement phase transition. Irregular behavior of this distribution
as a function of bombarding energy has been recently suggested~\cite{Iva10} as a signature of the QGP formation.

The $\pi^-$ rapidity distribution for the 10.7 AGeV central Au+Au collision is shown in Fig.~\ref{fig22}.
In the case of the EoS--PT, a reasonable agreement with experimental data can be achieved only if we
assume a freeze-out time smaller than for protons\,\footnote
{
Additional calculations show that freeze--out times for pions and nucleons become closer to each other
if one assumes smaller values of excluded volume for thermal pions as compared to nucleons.
}.
Again, calculations with the phase transition predict a broader distribution than those for the EoS--HG.
However, differences in shapes of pion rapidity distributions are not so strong as for protons.
It is worth to note that our results on hadronic spectra and collective flows (see next section) are rather
preliminary. First, they are obtained by using the simplest option of isochronous freeze-out.
Second, the resonance decays are considered rather schematically i.e.
the width of resonances was neglected and multi-particles decays were included only approximately.

\section{Collective flows}

       \begin{figure*}[htb!]
          \centerline{\includegraphics[width=0.9\textwidth]{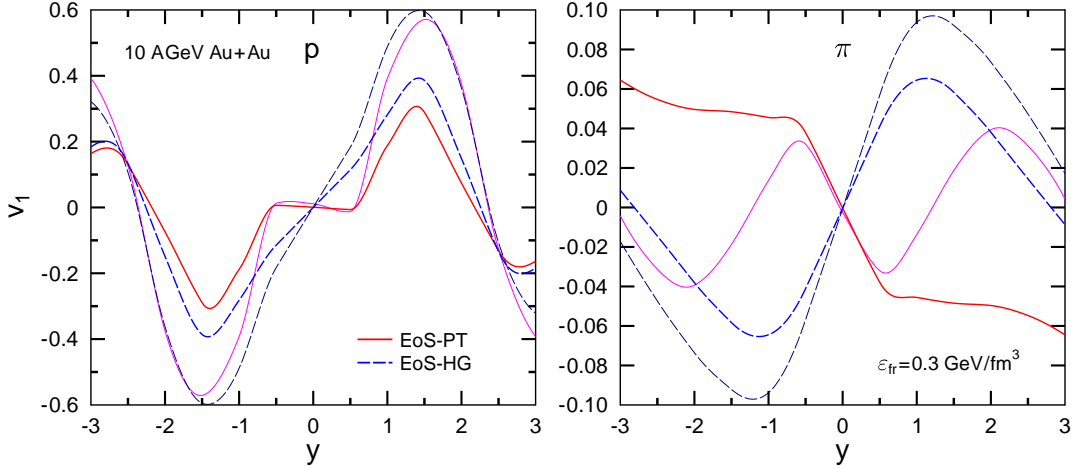}}
        \caption{(Color online)
        Directed flows of protons (left panel) and  charged pions (right panel)
        as functions of c.m. rapidity in 10 AGeV Au+Au collisions. Solid (dashed) lines are calculated with
        EoS--PT (EoS--HG). Thick and thin lines correspond to $b=4$ and $7$ fm, respectively.}
        \label{fig23}
        \end{figure*}
The momentum anisotropy considered in Sec.~\ref{moman} is not measurable directly and can be used only for
qualitative analysis of collective flows in nuclear collisions. On the other hand, theoretical and experimental studies
of the directed ($v_1$)and elliptic ($v_2$) flows became very popular in recent years (see e.g.~\cite{Sto05}).
In particular, this was initiated by the prediction~\cite{Ris95a} of a ''softest point collapse'' of the nucleon directed
flow at \mbox{$\ela\sim 8$ AGeV}. In this section we present our results for Au+Au and Pb+Pb  collisions
at $\ela=10$ and 40 AGeV. These results are obtained by using formulas of Sec.~\ref{specf}.

       \begin{figure*}[htb!]
       \vspace*{2mm}
          \centerline{\includegraphics[width=0.55\textwidth]{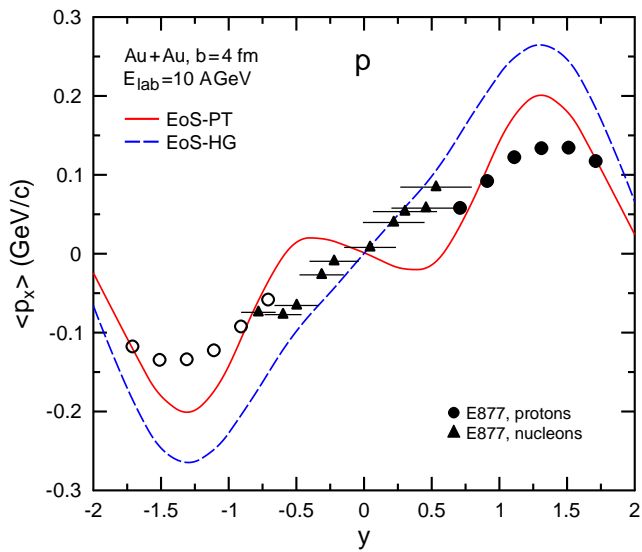}}
        \caption{(Color online)
        Mean directed transverse momentum of protons as functions of c.m. rapidity in 10 AGeV Au+Au collisions ($b=4$ fm).
        Solid and dashed lines are calculated with
        EoS--PT and EoS--HG, respectively. Experimental data~\cite{Bar97} for protons (full dots) and nucleons (full triangles)
        correspond to 10.1 AGeV midcentral Au+Au collisions. Circles are obtained by reflection.}
        \label{fig24}
        \end{figure*}
Figure~\ref{fig23} shows the model predictions for the proton and pion directed flows in 10 AGeV Au+Au collisions.
As before, we compare the results obtained for the \mbox{EoS--PT} and \mbox{EoS--HG}.
To demonstrate sensitivity to the centrality, we present results for two values
of the impact parameter, $b=4$ and 7 fm. In both cases it is assumed that freeze--out occurs when the energy density
in the central cell becomes lower than \mbox{$\varepsilon_{\rm fr}=0.3$ GeV/fm$^3$}.
The model predicts a~strong reduction of $|v_1^{(p)}|$ at small $|y|$ in transition from the EoS--HG to EoS--PT. One can see a noticeable
difference of slopes $dv_1/dy$ between two EoS\hspace*{0.5pt}s in a central rapidity region. In addition, negative slopes (antiflow)
of the pion and proton $v_1$ are predicted in the calculation with the EOS--PT. However, this effect is stronger for pions.
Our analysis shows that the antiflow in this case is formed mainly due to the contribution of thermal pions.
Note, that slopes \mbox{$dv_1^{(\pi)}/dy$} at \mbox{$|y|\lesssim 0.5$} only slightly depend on $b$.

       \begin{figure*}[htb!]
       \vspace*{2mm}
          \centerline{\includegraphics[width=1\textwidth]{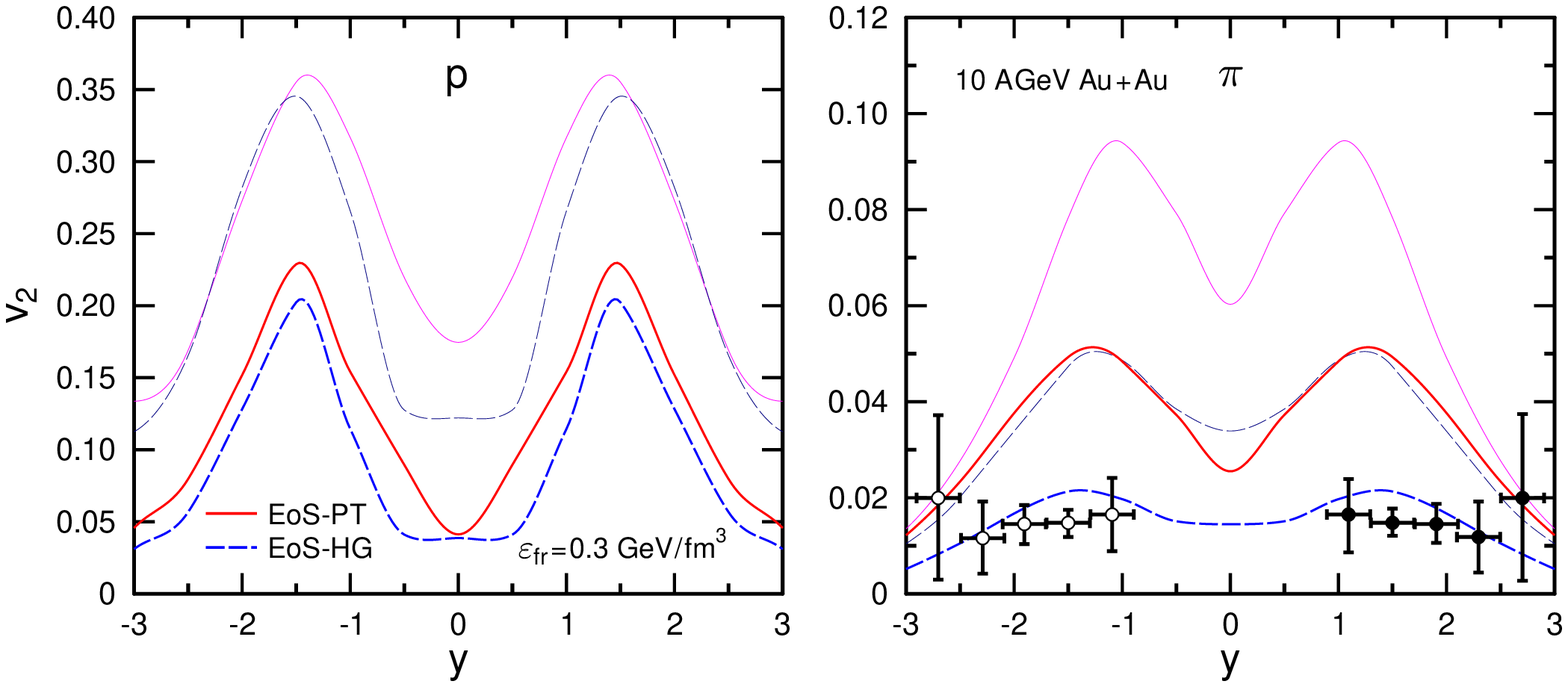}}
       \vspace*{-5mm}
        \caption{(Color online)
        Elliptic flows of protons (left panel) and charged pions (right panel)
        as functions of rapidity in 10 AGeV Au+Au collisions. Solid (dashed) lines are calculated with
        EoS--PT (EoS--HG). Thick and thin lines correspond to
        $b=4$ and $7$ fm, respectively. Full dots are experimental data~\cite{Fil99} for negative pions in 10.6 AGeV midcentral
        Au+Au collisions. Circles are obtained by reflection.}
        \label{fig25}
        \end{figure*}
Unfortunately, the rapidity dependence of directed flows is not measured at AGS bombarding energies. On the other hand, there are
measurements~\cite{Bar97} of a similar quantity, the so--called ''sideflow'' \mbox{$<p_{\hsp x}>$}, which is defined as a mean transverse momentum
(per particle) projected on the reaction plane. Below we calculate this quantity by using the r.h.s.~of~\re{eqv2} with the replacement
\mbox{$\cos{(2\phi)}\to p_{\hsp x}=p_{\hsp T}\cos{\phi}$}. Figure~\ref{fig24} shows the results for proton and pion sideflows in 10 AGeV Au+Au
collision with $b=4$ fm. On can see, that the calculation with the EOS--PT predicts
a back bending in the central bin, $|y|\lesssim 0.5$\hsp , which is not seen in data. On the other hand,
the observed maxima of \mbox{$|<p_{\hsp x}>|$} are stronger overestimated in the hadronic scenario.


     \begin{figure*}[htb!]
          \centerline{\includegraphics[width=\textwidth]{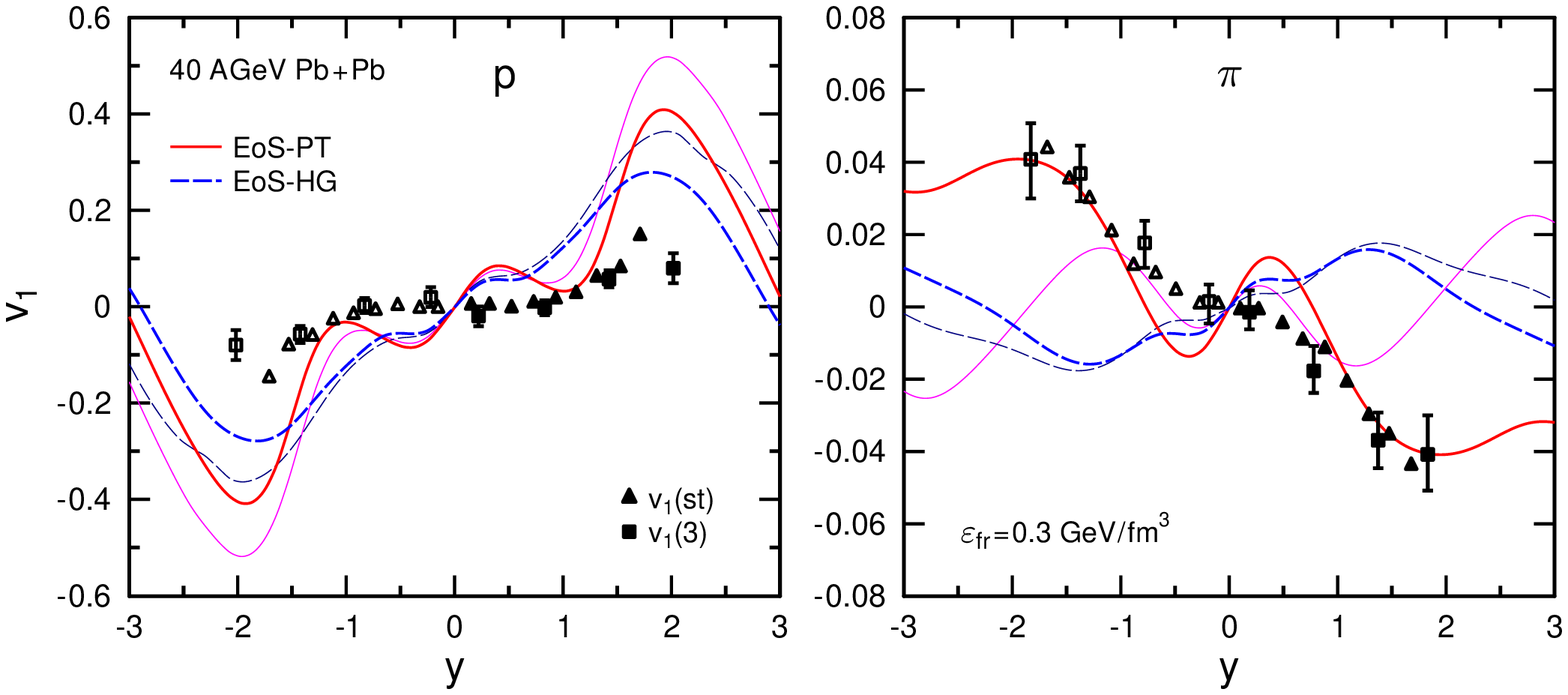}}
        \caption{(Color online)
        Directed flows of protons (left panel) and  charged pions (right panel)
        as functions of rapidity in 40 AGeV Pb+Pb collisions. Solid (dashed) lines are calculated with
        EoS--PT (EoS--HG). Thick and thin lines correspond to $b=4$ and $5.6$ fm, respectively.
        Full symbols are measured experimental data~\cite{Alt03}, while the open ones are
        obtained by reflection.}
        \label{fig26}
        \end{figure*}
Figure~\ref{fig25} represents proton and charged pion elliptic flows for the same reaction as in Figs.~\ref{fig23}--\ref{fig24}.
In agreement with discussion in Sec.~\ref{moman}, the deconfinement phase transition leads to enhancement of elliptic flows
as compared to the purely hadronic scenario. The relative increase of $v_2$ is larger for pions.
As before, we consider two values of the impact parameter, \mbox{$b=4$} and 7 fm. Experimental data exist only for $v_2^{(\pi)}$.
They are better reproduced with the EOS--HG. However, strong sensitivity of results to the choice of centrality and
possible influence of dissipative effects (see below) do not permit drawing a more definite conclusion.

       \begin{figure*}[htb!]
          \centerline{\includegraphics[width=\textwidth]{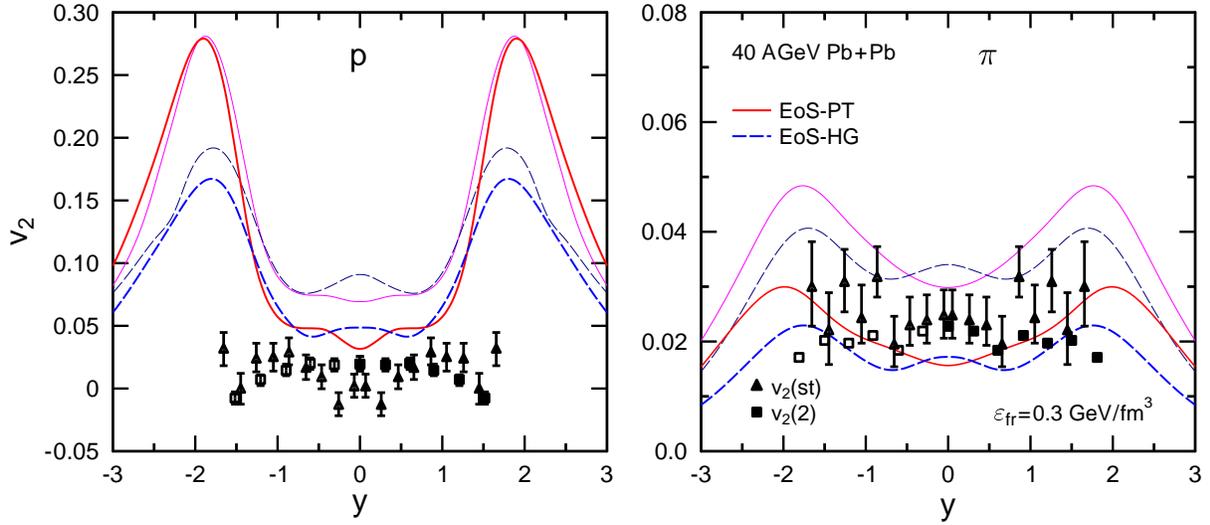}}
        \caption{(Color online)
        Same as Fig.~\ref{fig26}, but for
        elliptic flows of protons (left panel) and charged pions (right panel).
        }
        \label{fig27}
        \end{figure*}
To check the sensitivity of collective flows to the bombarding energy, we have performed calculations for
Pb+Pb collisions at $\ela=40$ AGeV. The results are shown in Figs.~\ref{fig26}--\ref{fig27}.
As compared to the AGS energy (see Fig.~\ref{fig23}) the proton directed flow is not so sensitive to the EoS,
at least in the central rapidity region. As one can see from Fig.~\ref{fig26}, the
calculations with the EoS--PT again predicts formation of the pion antiflow.
In this case experimental data are in qualitative agreement with our calculation while the EoS--HG leads
to the opposite (positive) slope of the directed flow.

Similar conclusion about smaller sensitivity of collective flows to the phase transition at SPS energies may
be drawn from Fig.~\ref{fig27}. Here we show the rapidity dependence of proton and pion elliptic flows for
40 AGeV Pb+Pb collisions. Note that experimental values of~$v_2$ strongly depend on the methods used for
their determination. One can see that the model overestimates the proton data at large c.m. rapidities.
As discussed in Ref.~\cite{Iva09a}, more definite conclusions can be made only after dissipative
and freeze--out effects will be carefully evaluated.

\section{Conclusions and outlook}

We have developed a (3+1) dimensional ideal hydrodynamical model to study heavy--ion collisions in the
energy range 1--160 AGeV. The sensitivity of collective flow observables and particle spectra
to the first--order deconfinement phase transition have been investigated. Our analysis shows that
maximal values of energy- and baryon densities in the central box of the colliding system are
significantly larger if this phase transition occurs at some intermediate stage of a heavy-ion collision.
As compared with the purely hadronic scenario, the calculations
with the deconfinement phase transition predict broadening of proton rapidity distributions,
enhancement of elliptic flows and appearance of directed antiflow in the central rapidity region.
It it shown that the collective flow parameters are especially sensitive to
EoS at \mbox{$\ela \simeq 10$ AGeV}. Our calculations with the EoS--PT predict that at such energies
the system spends the longest time in the mixed phase. Similar conclusion that this energy domain is optimal
for searching signatures of the deconfinement phase transition in nuclear collisions
has been made earlier in Ref.~\cite{Ars07}.

Our analysis does not allow us to decide which EoS, with or without the phase transition, gives better
agreement with experimental data. One should bear in mind that our
results are obtained by assuming a rather simple isochronous freeze-out and neglecting dissipative effects.
To detect clear signatures of the deconfinement transition in nuclear collisions, new detailed measurements
of collective flow observables, especially in the central rapidity region, would be needed in
NICA and FAIR experiments. In the future we plan to generalize our model by introducing dissipative terms
and a more realistic description of freeze-out process.

\begin{acknowledgments}
The authors thank Yu.B.~Ivanov for giving the results of the three--fluid model,
D.H.~Risch\-ke for providing us the one--fluid 3D code,
\mbox{M.I.~Gorenstein}, \mbox{P.~Huovinen}, \mbox{H.~Niemi},  D.Yu.~Pe\-re\-ssounko,
and V.D.~Toneev for useful discussions. The computational resources were
provided by the Center for Scientific Computing (Goethe University, Frankfurt am Main)
and by the Grid Computing Center (the Kurchatov Institute, Moscow).
This work was supported in part by the DFG \mbox{grant 436 RUS 113/957/0--1},
the Helmholtz International Center for FAIR~(Germany) and
the grants RFBR 09--02--91331 and NSH--7235.2010.2 (Russia).
\end{acknowledgments}



\begin{thebibliography}{00}
\bibitem{Baz09}
        A. Bazavov \textit{et al.}, Phys. Rev. D \textbf{80} 014504 (2009).

\bibitem{Aok09}
        Y. Aoki \textit{et al.}, JHEP \textbf{0906} 088 (2009).

\bibitem{Ada05}
        STAR Collaboration, J. Adams \textit{et al.},
        Nucl. Phys. A \textbf{757}, 192 (2005);\\
        PHENIX Collaboration, K. Adcox \textit{et al.},
        Nucl. Phys. A \textbf{757}, 164 (2005);\\
        PHOBOS Collaboration, B.B. Back \textit{et al.},
        Nucl. Phys. A \textbf{757}, 28 (2005).

\bibitem{Alf98}
        M.G. Alford, K. Rajagopal, and F. Wilzcek, Phys. Lett. B \textbf{422} 247 (1998).

\bibitem{Sca01}
        O. Scavenius, A. Mocsy, I.N. Mishustin, and D.H. Rischke,
        Phys. Rev. C \textbf{64} 045202 (2001).

\bibitem{Ste04}
        M.A. Stephanov, Prog. Theor. Phys. Suppl. \textbf{153}, 139 (2004)
        [Int. J. Mod. Phys. A \textbf{20}, 4387 (2005)].

\bibitem{Rit06}
        H.G. Ritter, PoS CPOD2006, 015 (2006);\\
        G. Odyniec, J. Phys. G: Nucl. Part. Phys. \textbf{37} 094028 (2010).

\bibitem{Nic11}
        NICA white paper, http://theor.jinr.ru/twiki-cgi/view/NICA/NICAWhitePaper.

\bibitem{Fai06}
        [FAIR] An International Accelerator Facility for Beams of Ions and Antiprotons, Conceptual
        Design Report, http://www.gsi.de/GSI-Future/cdr/.

\bibitem{Lan53}
        L.D. Landau, Izv. Akad. Nauk Ser. Fiz. \textbf{17}, 5
        (1953); in \textit{Collected Papers of L.D. Landau},
        Gordon and Breach, New York, 1965, p.~665.

\bibitem{Bjo83}
        J.D. Bjorken, Phys. Rev. D \textbf{27}, 140 (1983).

\bibitem{Mis83}
        I.N. Mishustin and L.M. Satarov,
        Yad.~Fiz. \textbf{37}, 894 (1983)
        [\mbox{Sov.~J.~Nucl.~Phys.}~\textbf{37}, 532 (1983)].

\bibitem{Bla87}
        J.P. Blaizot and J.Y. Ollitrault,
        Phys. Rev. D \textbf{36}, 916 (1987).

\bibitem{Esk98}
        K.J. Eskola, K. Kajantie, and P.V. Ruuskanen,
        Eur. Phys. J. C \textbf{1}, 627 (1998).

\bibitem{Moh03}
        B. Mohanty and J. Alam, Phys. Rev. C \textbf{68},
        064903 (2003).

\bibitem{Sat07}
        \mbox{L.M. Satarov, I.N. Mishustin, A.V. Merdeev, and H. St\"ocker,
        Phys. Rev. C \textbf{75}, 024903 (2007);}\\
        Yad. Fiz. \textbf{70}, 1822 (2007)
        \mbox{[Phys. Atom. Nucl. \textbf{70}, 1773 (2007)\hsp]\hsp}.

\bibitem{Kol99}
         P.F. Kolb, J. Sollfrank, and U. Heinz, Phys. Lett.
         B~\textbf{459}, 667 (1999).

\bibitem{Bas00}
        S.A. Bass and A. Dumitru, Phys. Rev. C  \textbf{61}, 064909 (2000).

\bibitem{Per00}
         D.Yu. Peressounko and Yu.E. Pokrovsky, Nucl. Phys. A~\textbf{669}, 196 (2000).

\bibitem{Tea01}
        D. Teaney, J. Lauret, and E.V. Shuryak,
        Phys. Rev. Lett. \textbf{86}, 4783 (2001);
        \mbox{nucl--th/0110037}.

\bibitem{Sol97}
        J. Sollfrank \textit{et al.}
        Phys. Rev. C~\textbf{55}, 392 (1997).

\bibitem{Non00}
        C. Nonaka, E. Honda, and S. Muroya, Eur. J. Phys. C, \textbf{17}, 663 (2000).

\bibitem{Agu02}
        C.E. Aguiar, Y. Hama, T. Kodama, and T. Osada,
        Nucl. Phys. A \textbf{698}, 639 (2002).

\bibitem{Hir02}
        T. Hirano and K. Tsuda, Phys. Rev. C~\textbf{66}, 054905 (2002).

\bibitem{Boz09a}
        P. Bozek and I. Wyskiel, Phys. Rev. C \textbf{79}, 044916 (2009).

\bibitem{Sch10}
        B. Schenke, S. Jeon, and C. Gale, Phys. Rev. C \textbf{82}, 014903 (2010).

\bibitem{Son08}
        H. Song and U. Heinz, Phys. Rev. C \textbf{77}, 064901 (2008).

\bibitem{Luz08}
        M. Luzum and P. Romatschke, Phys. Rev. C \textbf{78}, 034915 (2008);
        \textbf{79}, 039903(E) (2009).


\bibitem{Ams75}
        A.A. Amsden, G.F. Bertsch, F.H. Harlow, and J.R. Nix,
        Phys. Rev. Lett.~\textbf{35}, 905 (1975).

\bibitem{Sto79}
        H. St\"ocker, J.A. Maruhn, and W. Greiner,
        Z. Phys. A \textbf{290}, 297 (1979).

\bibitem{Ros81}
        A.S. Roshal and V.N. Russkikh,
        Yad. Fiz. \textbf{33}, 1520 (1981)
        [Sov. J. Nucl. Phys. \textbf{33}, 817 (1982)\hsp]\hsp.

\bibitem{Bra94}
        L. Bravina, L.P. Csernai, P. Le'vai, and D. Strottman,
        Phys. Rev. C \textbf{50}, 2161 (1994).

\bibitem{Ris95a}
        D.H. Rischke, Y. P\"urs\"un, J.A. Maruhn, H. St\"ocker,
        and W. Greiner, Heavy Ion Phys. \textbf{1}, 309~(1995).

\bibitem{Pae01}
        K. Paech, M. Reiter, A. Dumitru, H. St\"ocker, and W. Greiner,
        Nucl. Phys. A \textbf{681}, 41c (2001).

\bibitem{Ams78}
        A.A. Amsden, A.S. Goldhaber, F.H. Harlow, and J.R. Nix,
        Phys. Rev.~C~\textbf{17}, 2080 (1978).

\bibitem{Cla86}
        R.B. Clare and D. Strottman,
        Phys. Rep. \textbf{141}, 178 (1986).

\bibitem{Mis88}
    I.N.~Mishustin, V.N.~Russkikh, and L.M.~Satarov,
        \mbox{Yad.~Fiz. \textbf{48}, 711 (1988)}\\~
        [Sov.~J.~Nucl.~Phys. \textbf{48}, 454 (1988)];
        Nucl. Phys. A~\textbf{494}, 595 (1989).

\bibitem{Kat93}
        U. Katscher, D.H. Rischke, J.A. Maruhn, W. Greiner,
        I.N. Mishustin, and L.M.~Satarov,\\Z.~Phys. A ~\textbf{346}, 209 (1993).

\bibitem{Bra00}
        J. Brachmann \textit{et al.}, Phys. Rev. C~\textbf{61}, 024909 (2000).

\bibitem{Iva06}
        Yu.B. Ivanov, V.N. Russkikh, and V.D. Toneev,
        Phys. Rev. C~\textbf{73}, 044904 (2006).

\bibitem{Coo74}
        F. Cooper and G. Frye, Phys. Rev. D \textbf{10}, 186 (1974).

\bibitem{Bug96}
        K. Bugaev, Nucl. Phys. A \textbf{606}, 559 (1996).

\bibitem{Hun98}
        C. M. Hung and E. Shuryak, Phys. Rev. C~\textbf{57}, 1891 (1998).


\bibitem{Non07}
        C. Nonaka and S.A. Bass, Phys. Rev. C \textbf{75}, 014902 (2007).

\bibitem{Hir08}
        T. Hirano, U.W. Heinz, D. Kharzeev, R. Lacey, and Y. Nara,
        Phys. Rev. C \textbf{77}, 044909 (2008).

\bibitem{Pet08}
        H. Petersen, J. Steinheimer, G. Bereau, M. Bleicher, and H. St\"ocker,
        Phys. Rev. C \textbf{78}, 044901 (2008).

\bibitem{Sat09}
        \mbox{L.M. Satarov, M.N. Dmitriev, and I.N. Mishustin,
        Yad. Fiz. \textbf{72}, 1444 (2009)}
        [Phys.~Atom. Nucl. \textbf{72}, 1390 (2009)\hsp]\hsp.

\bibitem{Lan87}
         L.D. Landau and E.M. Lifshitz, \textit{Fluid
         Mechanics}, Pergamon Press, 1987.

\bibitem{Bor73}
        J.P. Boris and D.L. Book, J. Comp. Phys. \textbf{11}, 38 (1973).

\bibitem{Ris95b}
        D.H. Rischke, S. Bernard, and J.A. Maruhn,
        Nucl. Phys. A \textbf{595}, 346 (1995).

\bibitem{Lan80}
        L.D. Landau and E.M. Lifshitz, \textit{Statistical Physics},
        Pergamon press, 1980.

\bibitem{Nak10} K. Nakamura \textit{et al.} (Particle Data Group),
        J. Phys. G \textbf{37}, 075021 (2010).

\bibitem{Oll92}
        J.-Y. Ollitrault, Phys. Rev. D \textbf{46}, 229 (1992).

\bibitem{Kol00}
        P.F. Kolb, J. Sollfrank, and U. Heinz,
        Phys. Rev. C \textbf{62}, 054909 (2000).

\bibitem{Iva09a}
        Yu.B. Ivanov, I.N. Mishustin, V.N. Russkikh, and L.M. Satarov,
        Phys. Rev. C \textbf{80}, 064904 (2009).

\bibitem{Pet10}
        H. Petersen and M. Bleicher,  Phys. Rev. C \textbf{81},
        044906 (2010).

\bibitem{Mis91}
        I.N.~Mishustin, V.N.~Russkikh, and L.M.~Satarov,
        \mbox{Yad.~Fiz. \textbf{54}, 429 (1991)}\\~
        [Sov.~J.~Nucl.~Phys. \textbf{54}, 260 (1991)].

\bibitem{Boz09}
        P. Bozek, Phys. Rev. C \textbf{79}, 054901 (2009).

\bibitem{Iva09b}
        Yu.B. Ivanov, private communication.

\bibitem{Ton04}
        V.D. Toneev \textit{et al.},
        Eur. Phys. J. C \textbf{32}, 399 (2004).

\bibitem{Ahl98}
        L. Ahle \textit{et al.} (E802 Collaboration),
        Phys. Rev. C \textbf{57}, R466 (1998).

\bibitem{Bac01}
        B.B. Back \textit{et al.} (E917 Collaboration),
        Phys. Rev. Lett.~\textbf{86}, 1970 (2001).

\bibitem{Bar00}
        J. Barrette \textit{et al.} (E877 Collaboration),
        Phys. Rev. C~\textbf{62}, 024901 (2000).


\bibitem{Bra97}
        J. Brachmann, A. Dumitru, J.A. Maruhn, H. St\"ocker, W. Greiner,
        and D.H. Rischke, Nucl.~Phys. A \textbf{616}, 391 (1997).

\bibitem{Iva10}
        Yu.B. Ivanov, Phys. Lett. B \textbf{690} 358 (2010).


\bibitem{Kla03}
        J.L. Klay \textit{et al.} (E895 Collaboration),
        Phys. Rev. C~\textbf{68}, 054905 (2003).


\bibitem{Sto05}
        H. St\"ocker, Nucl. Phys. A~\textbf{750}, 121 (2005).

\bibitem{Bar97}
        J. Barrette \textit{et al.} (E877 Collaboration),
        Phys. Rev. C~\textbf{56}, 3254 (1997).

\bibitem{Fil99}
        K. Filimonov \textit{et al.} (E877 Collaboration),
        Nucl. Phys. A \textbf{661}, 198 (1999).


\bibitem{Alt03}
        C. Alt \textit{et al.} (NA49 Collaboration),
        Phys. Rev. C~\textbf{68}, 034903 (2003).

\bibitem{Ars07}
        I. C. Arsene \textit{et al.},
        Phys. Rev. C~\textbf{75}, 034902 (2007).

\end{thebibliography}
\end{document}